\documentclass{article}
\usepackage{epsf}

\newcommand{\bookfig}[5]{
\begin{figure}\centering\fbox{\epsfysize=#5cm \epsfbox{#1}}
\caption[#2]{\small #4}\label{#3}
\end{figure}
}

\def\res{\;\raisebox{-2.5mm}{\epsfysize=2cm\epsfbox{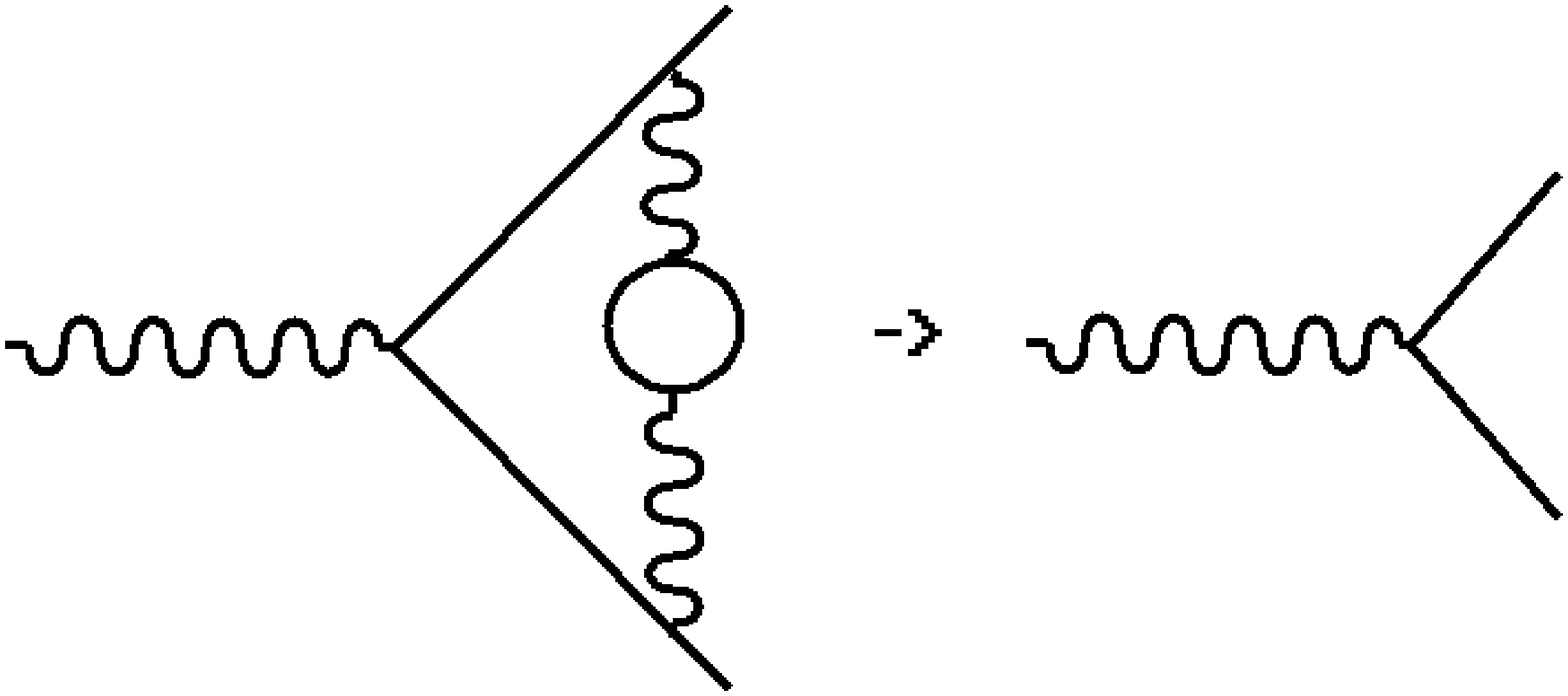}}\;}
\def\pica{\;\raisebox{-2.5mm}{\epsfysize=4cm\epsfbox{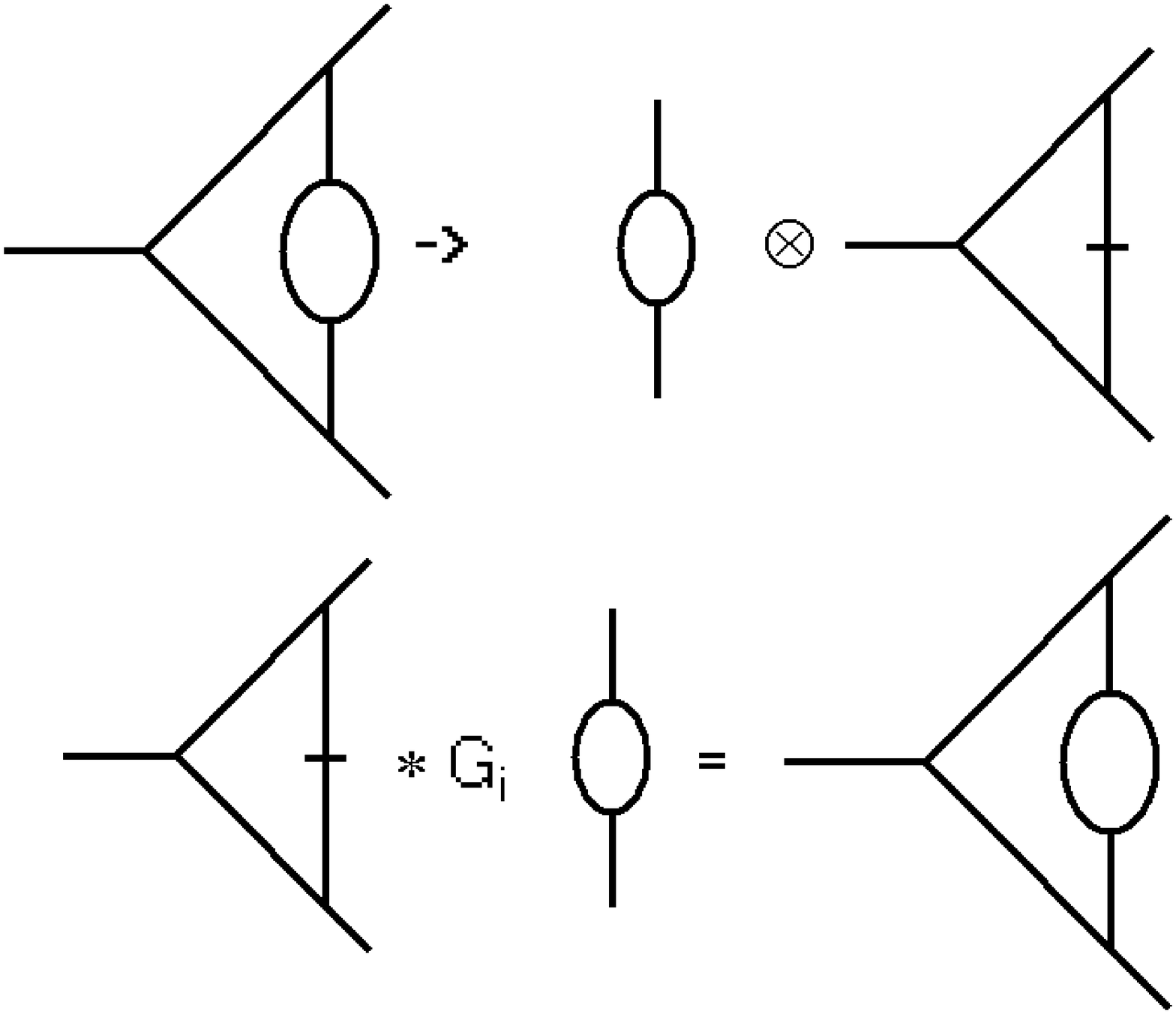}}\;}
\def\picb{\;\raisebox{-2.5mm}{\epsfysize=4cm\epsfbox{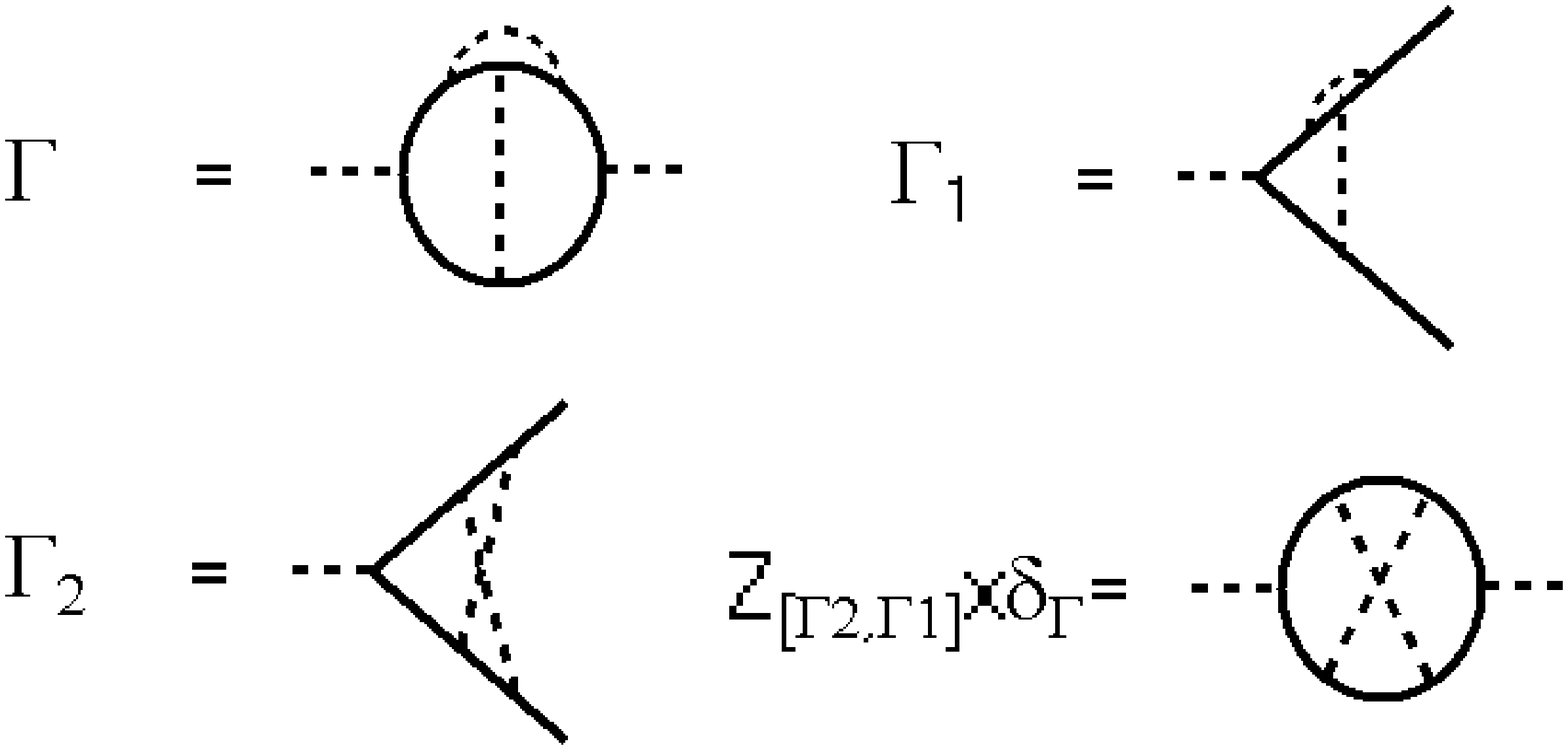}}\;}

\def\vv{\;\raisebox{-6mm}{\epsfysize=15mm\epsfbox{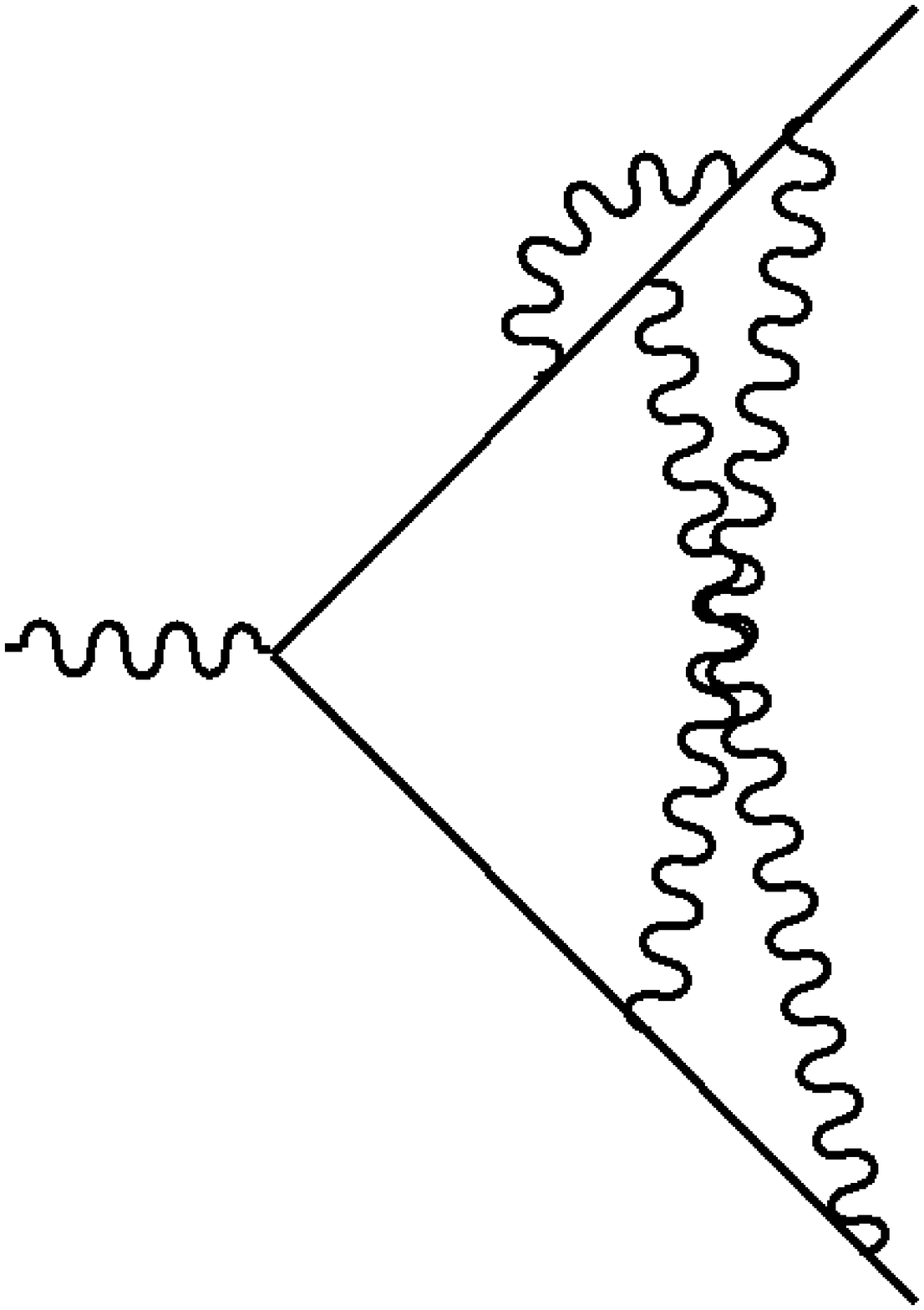}}\;}
\def\fo{\;\raisebox{-3mm}{\epsfysize=8mm\epsfbox{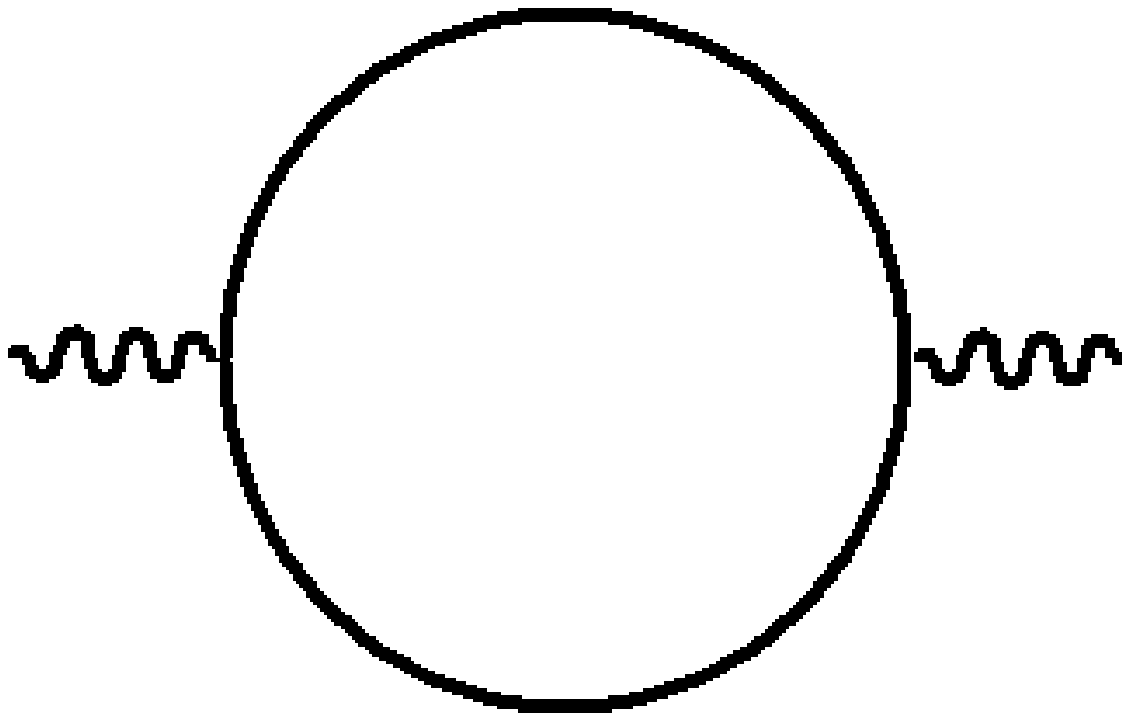}}\;}
\def\fs{\;\raisebox{-3mm}{\epsfysize=8mm\epsfbox{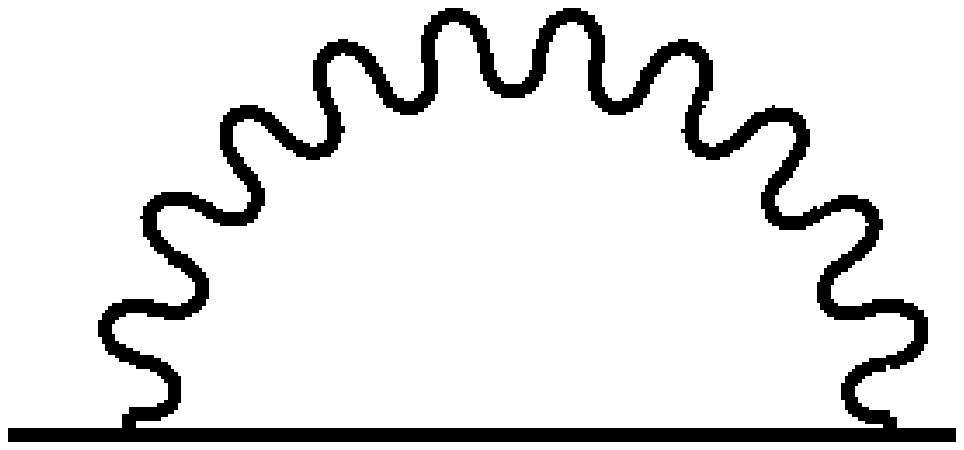}}\;}

\def\vvx{\;\raisebox{-6mm}{\epsfysize=15mm\epsfbox{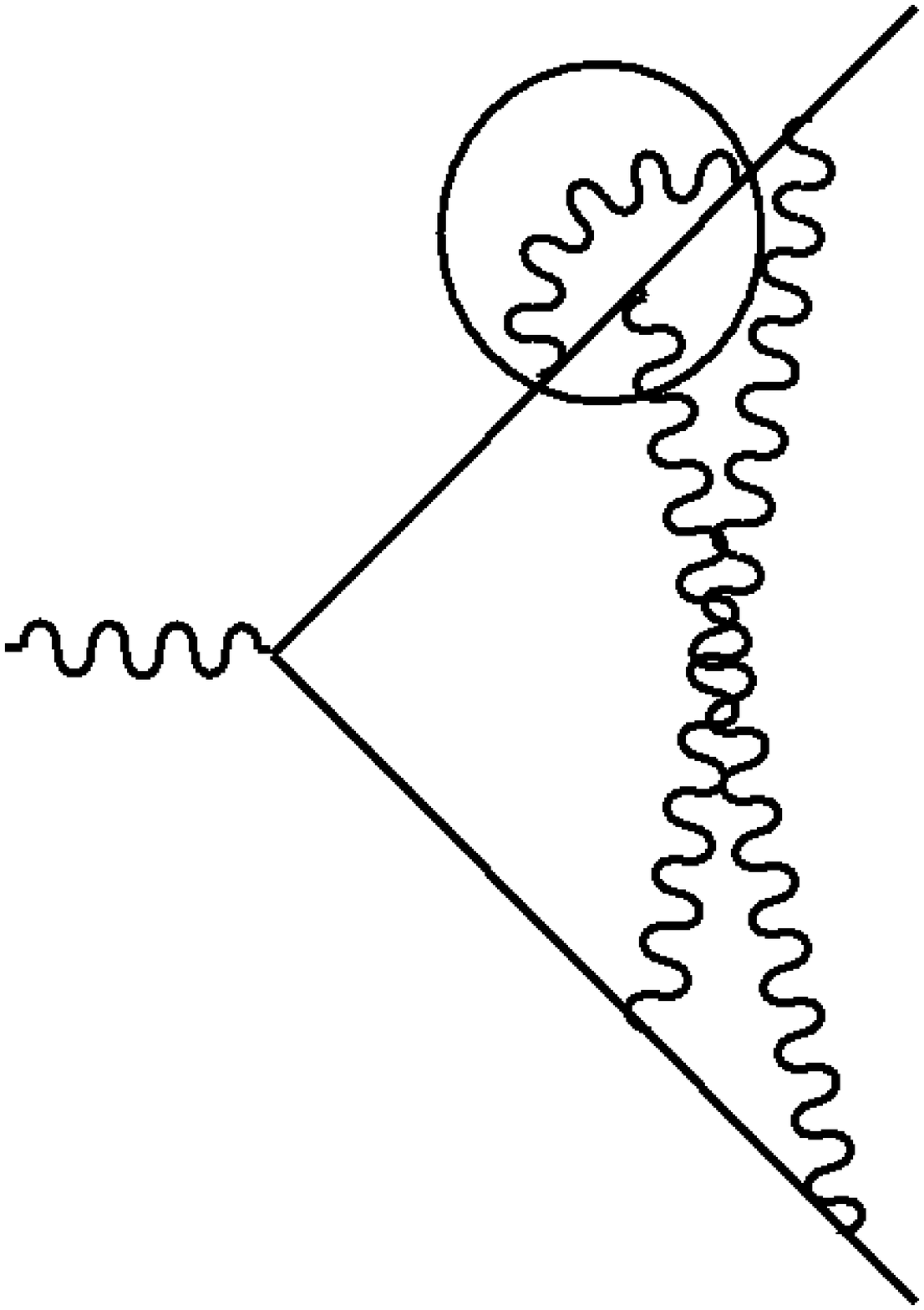}}\;}

\def\fx{\;\raisebox{-12mm}{\epsfysize=25mm\epsfbox{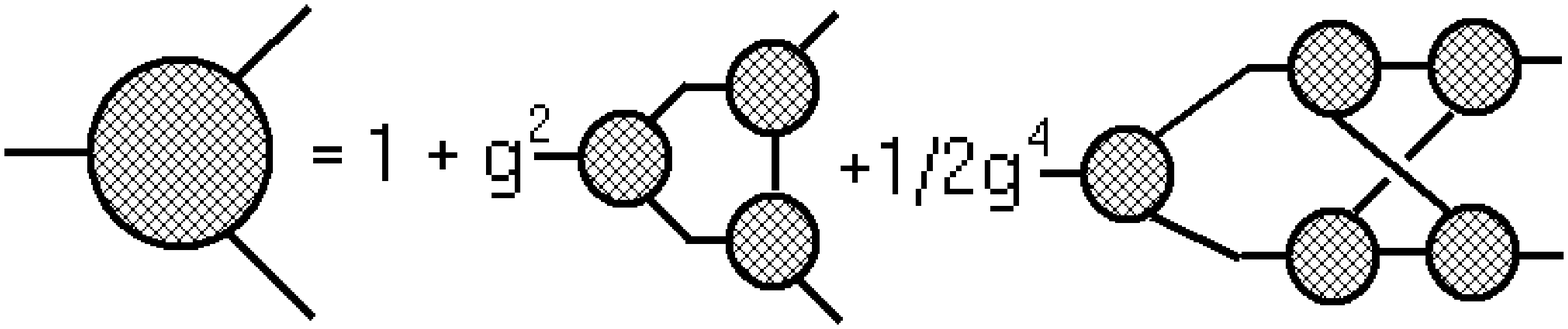}}\;}
\def\fy{\;\raisebox{-20mm}{\epsfysize=40mm\epsfbox{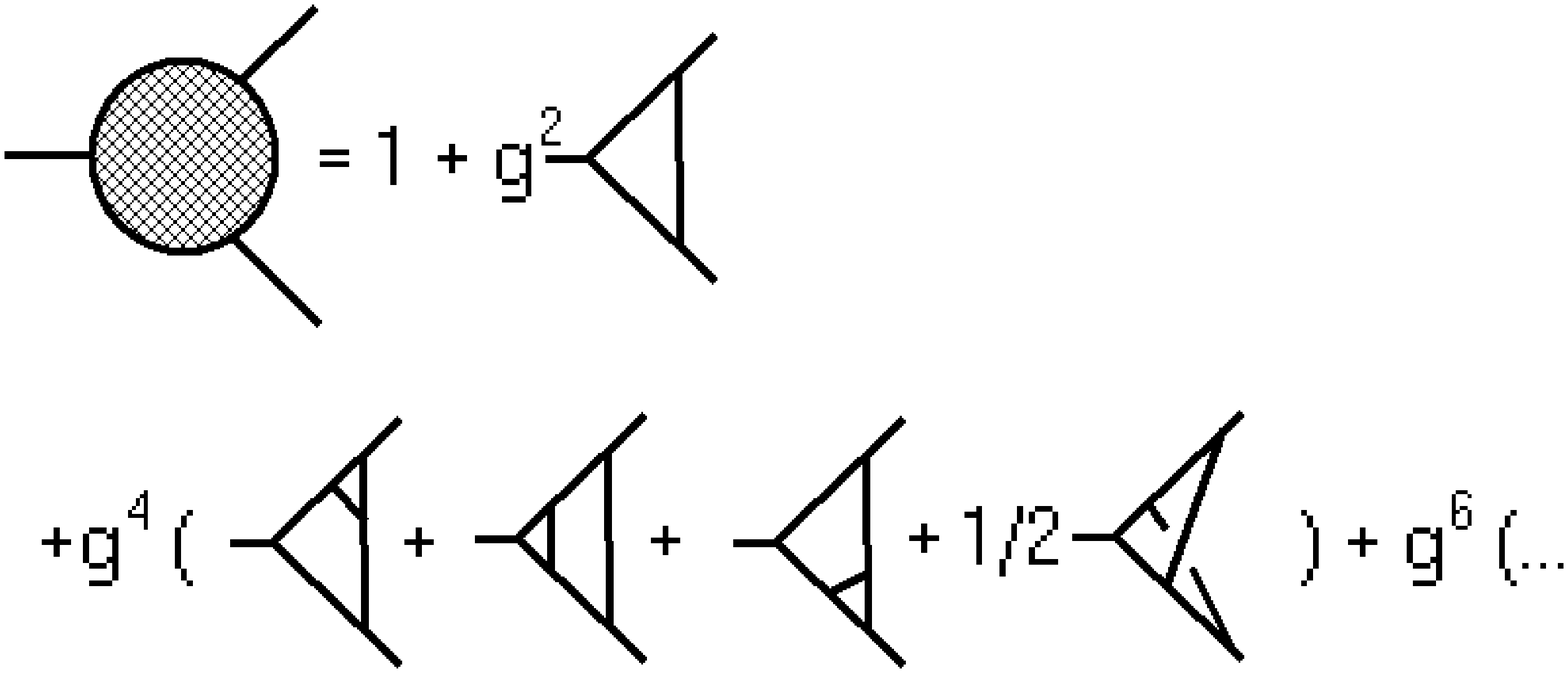}}\;}
\def\fz{\;\raisebox{-50mm}{\epsfysize=60mm\epsfbox{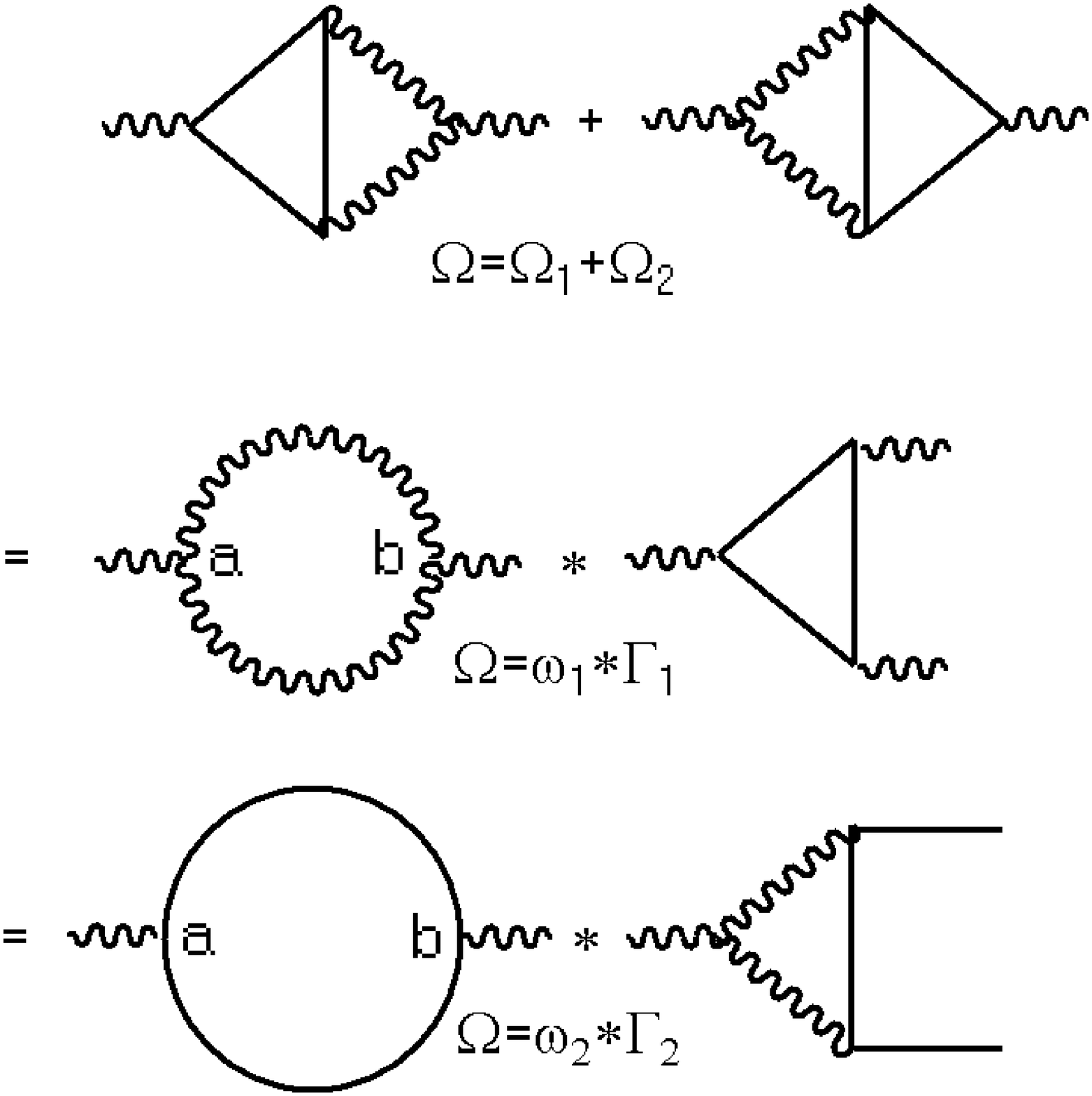}}\;}

\def\vs{\;\raisebox{-6mm}{\epsfysize=15mm\epsfbox{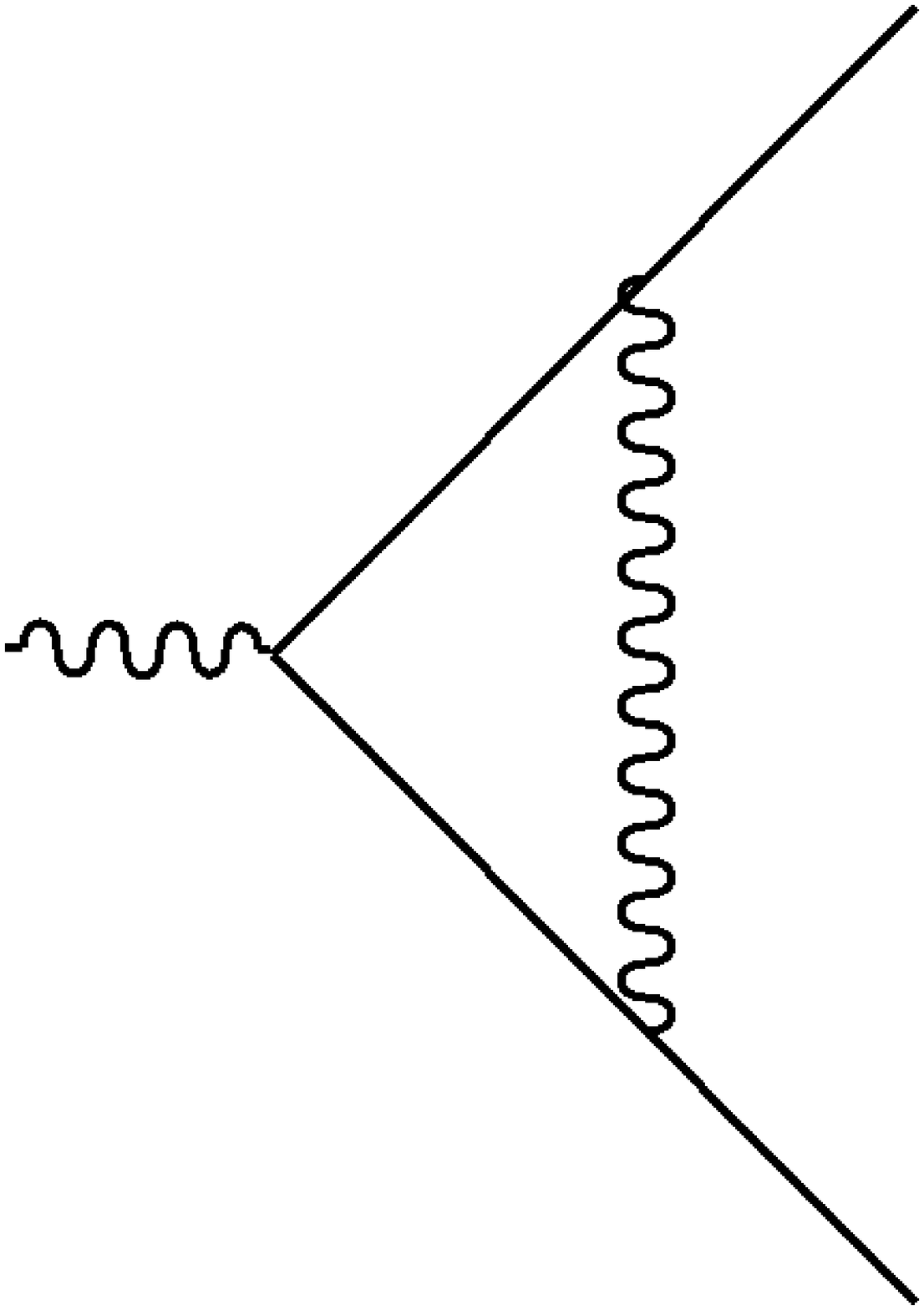}}\;}

\def\vc{\;\raisebox{-6mm}{\epsfysize=15mm\epsfbox{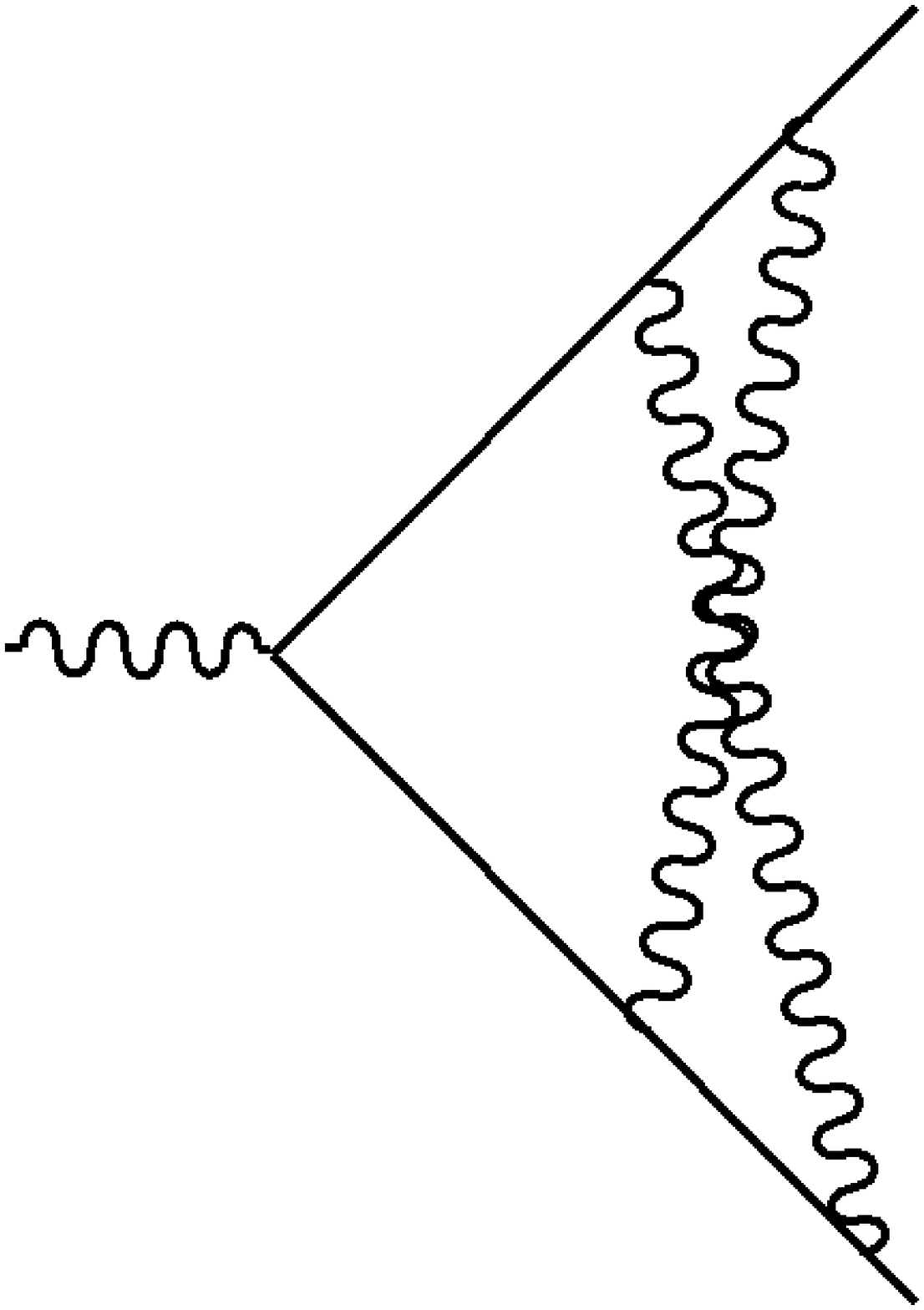}}\;}

\def\vvm{\;\raisebox{-6mm}{\epsfysize=15mm\epsfbox{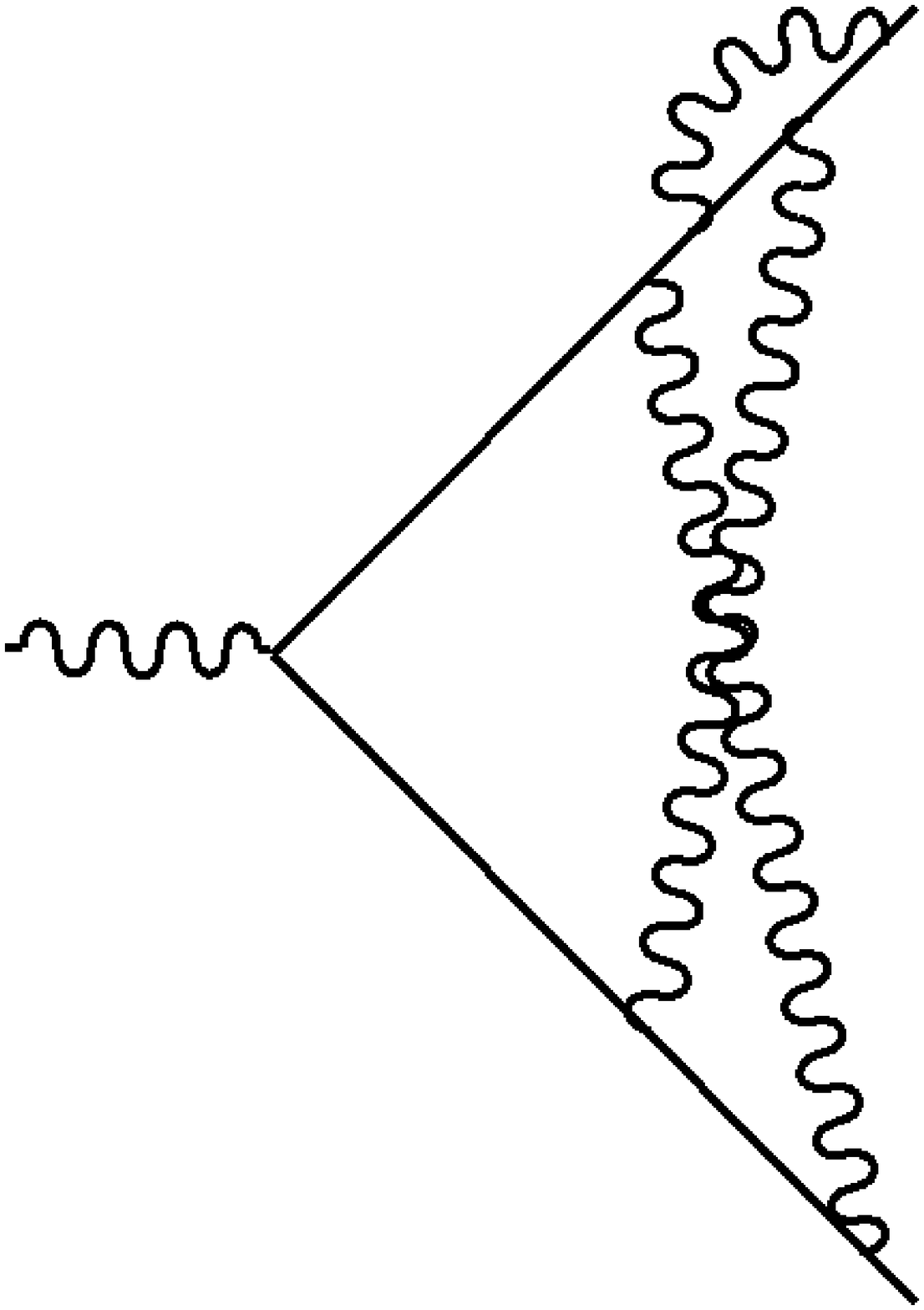}}\;}

\def\build#1_#2^#3{\mathrel{
\mathop{\kern 0pt#1}\limits_{#2}^{#3}}}

\def\ra{\rightarrow}

\def\fl{\forall}
\def\ify{\infty}

\def\pslash{p\!\!\!/}

\def\b{\beta}

\def\g{\gamma}

\def\t{\theta}
\def\ve{\varepsilon}

\font\tenbb=msbm10 \font\sevenbb=msbm7 \font\fivebb=msbm5
\newfam\bbfam
\textfont\bbfam=\tenbb \scriptfont\bbfam=\sevenbb
\scriptscriptfont\bbfam=\fivebb

\def\Hc{{\cal H}}

\catcode`\@=11
\def\displaylinesno #1{\displ@y\halign{
\hbox to\displaywidth{$\@lign\hfil\displaystyle##\hfil$}&
\llap{$##$}\crcr#1\crcr}}

\def\ldisplaylinesno #1{\displ@y\halign{
\hbox to\displaywidth{$\@lign\hfil\displaystyle##\hfil$}&
\kern-\displaywidth\rlap{$##$} \tabskip\displaywidth\crcr#1\crcr}}
\catcode`\@=12

\def\build#1_#2^#3{\mathrel{
\mathop{\kern 0pt#1}\limits_{#2}^{#3}}}

\newcommand{\be}{\begin{equation}}
\newcommand{\ee}{\end{equation}}
\newcommand{\bea}{\begin{eqnarray}}
\newcommand{\eea}{\end{eqnarray}}
\newcommand{\beas}{\begin{eqnarray*}}
\newcommand{\eeas}{\end{eqnarray*}}

\newtheorem{theorem}{Theorem}

\newtheorem{prop}[theorem]{Proposition}
\newtheorem{cor}[theorem]{Corollary}
\newtheorem{defn}[theorem]{Definition}

\font\tenbb=msbm10 \font\sevenbb=msbm7 \font\fivebb=msbm5
\newfam\bbfam
\textfont\bbfam=\tenbb \scriptfont\bbfam=\sevenbb
\scriptscriptfont\bbfam=\fivebb

\begin{document}
\title{Structures in Feynman Graphs -
Hopf Algebras and Symmetries\footnote{Talk given at the {\em
Dennisfest}, SUNY at Stony Brook, June 14-21 2001}}
\author{{Dirk Kreimer\thanks{dkreimer@bu.edu}}\\[5mm]{\bf\small
CNRS-IHES} {\small and} {\bf\small Center Math.\ Phys., Boston
U.}} \maketitle \vspace{-22em}\begin{flushright} {\small
BUCMP/02-02\\
hep-th/0202110}
\end{flushright}\vspace{18em}

\begin{abstract}
We review the combinatorial structure of perturbative quantum
field theory with emphasis given to the decomposition of graphs
into primitive ones. The consequences in terms of unique
factorization of Dyson--Schwinger equations into Euler products
are discussed.
\end{abstract}

\section*{Introduction}
The reputation of quantum field theory has always been mixed. As a
predictive theory, it is the best theory ever formulated. It has
been plagued by inconsistencies and conceptual flaws though, ever
since it was first spelled out. Roughly speaking, these
shortcomings come in two forms: order by order in the perturbation
theory short-distance singularities seemingly destroy the meaning
of the Feynman rules, and the predictive power of a perturbative
calculation. Elimination of this flaw in perturbative
renormalization works self-consistently, but this does not satisfy
a mathematician: without any guiding structure, renormalization
remained ill-reputed as a sole means to hide the infinities under
the carpet. Surely not a pillar on which the foundations of the
theory could rest. Eventually, recourse to extended objects,
avoiding the presence of point-like short-distance singularities,
seemed unavoidable. So far, this has not led to the advent of a
predictive theory replacing local quantum field theory.

The second shortcoming of perturbative quantum field theory was
its inability to make contact to non-perturbative approaches: a
demon, enabled to renormalize any loop order in arbitrary short
time indeed seems to be monstrous: we are confronted typically
with a series of finite numbers whose asymptotic behaviour defies
understanding, ie.~resummation so far. Singularities in the Borel
plane on the positive axis can be generated by renormalons, and by
instanton singularities \cite{Wightman}.

Part of these flaws gave way recently: the ugly duckling of
short-distance singularities and their elimination in perturbative
renormalization turned out to be a conceptual asset of the theory.

We will review these developments and put them into context, with
emphasis given to comment on future potential for progress beyond
perturbation theory. Much of what is reported here in the first
six sections  has been published elsewhere, or was, in much
greater detail, the content of a course in renormalization theory
recently given \cite{BU}. The final section then is devoted to
some new ideas.

\section{Lie- and Hopf algebra structures in a perturbative expansion}
\subsection{Motivation}
The structure of the perturbative expansion of a Quantum Field
Theory (QFT) is in many ways determined by the Hopf and Lie
algebra structures of Feynman graphs. Forest formulas originate
from the Hopf algebra structure, while notions like anomalous
dimensions and $\beta$-functions relate to the Lie algebra
structure. This allows for considerable simplifications in the
conceptual interpretation of renormalization theory. Indeed, the
identification with the Riemann--Hilbert problem allows to
summarize renormalization theory in a single line: find the
Birkhoff decomposition of a regularized but unrenormalized
physical parameter of interest. The positive part will be its
renormalized contribution (in a MS scheme), the negative part the
corresponding counterterm \cite{RHI,RHII}.

Nevertheless, this result, a direct consequence of the Hopf and
Lie algebra structure and of the existence of a group homomorphism
to diffeomorphism groups, does not exhaust the tools given at our
hand by these algebraic structures.

It indeed seems wise to start a consideration of quantum field
theory from the viewpoint of combinatorics and graph theory, a
viewpoint already mandated by 't Hooft and Veltman's famous
diagrammar \cite{diagrammar}. By its very definition, QFT will
ultimately reflect, in its short distance singularities and its
most notorious properties reflecting those, the structure of
spacetime at the infinitesimal small. In the lack of the ability
to perform experiments at essentially infinitely high energies,
the observables which arise from the presence of short distance
singularities are the only window we have towards that structure.

But then, is is desirable to rest the pillars of the foundations
of QFT on structures which are robust enough to accommodate the
unknown structure of the very small, and hence combinatorics is
certainly a good candidate down here.

From this viewpoint, it seems favorable to start from Feynman
diagrams, and try to derive the features which we hope to see in a
QFT from their combinatorial properties. The structure of the very
small might still be queerer than we think, and maybe even queerer
than we can think, and so any attempt to axiomatize or  construct
QFT from principles gained from experience with the not so very
small might ultimately turn out to be demanding more than Nature
is prepared to deliver. So we will set out to explore the
combinatorial structures behind a perturbative expansion, which,
as we will see, in itself provides the means to handle
short-distance singularities, and offers much in terms of a
conceptual analysis of QFT. The development of this combinatorial
viewpoints owns much to the efforts of practitioners of QFT, who
exposed it to the most cruel tests in radiative correction
calculations. One is left with awe when one studies in detail how
well perturbative QFT fares in such tests. None of the rigorous
approaches to QFT ever produced tools which contributed to the art
of radiative correction calculations significantly while the
combinatorial notions reported here build a rigorous  mathematical
background for the practice of QFT, and hopefully start to close a
gap between such practice of QFT and its mathematical foundations
which grew far too big in the last decades.

There are two basic operations on Feynman graphs which govern
their combinatorial structure, organize their contributions to a
chosen Green function as well as organize the process of
renormalization.

These two basic operations are the decomposition of a graph into
subgraphs, and the opposite operation, insertion of subgraphs into
a graph. While insertion of subgraphs is needed to generated the
formal series over graphs which provide a fixpoint for the
Dyson--Schwinger equation of a given Green function, decomposition
of graphs is necessary to achieve renormalization by counterterms
which are local expressions, polynomial in (derivatives of) fields
in the Lagrangian. Such a Lagrangian ${\cal L}$ is typically a
finite sum of monomials $${\cal L}=\sum_i {\cal M}_i,$$ where, for
example in a massive scalar theory with cubic interaction, we have
monomials ${\cal M}_1=1/2\;\partial_\mu\phi\partial^\mu\phi$,
${\cal M}_2=1/2\;m^2\phi^2$, ${\cal M}_3=g/6\;\phi^3$. Each such
monomial ${\cal M}_i$ can obtain a $Z$-factor ${\cal M}_i\to Z_i
{\cal M}_i$ to absorb short-distance singularities. Feynman graphs
arise when we expand in terms of a weak coupling $g$. The
$Z$-factors provide invertible series in $g$, their constant term
is unity. The theory is typically calculated using some regulator.
$Z$-factors are arranged such that they eliminate all divergences
so that the regulator can be switched off eventually. As always,
absorbing singularities allows for choice of the remaining finite
part, which gives rise to the various renormalization schemes used
in practice.

 Let us have a first
look at a Feynman graph and the roles these operations play.
Consider a three-loop vertex-correction $\Gamma$, this time in QED
in four dimensions, with the usual identification of wavy lines
with photons and straight lines with fermions {\Large
$$\Gamma=\vv.$$} This graph $\Gamma$ consists of twelve edges and
seven vertices. We denote by $\Gamma^{[0]}$ the set of vertices,
and by $\Gamma^{[1]}$ the set of edges. There are three external
edges which have an open end. They are just a reminder of the
meaning a physicist gives to such a graph: it is a contribution to
the probability amplitude of a scattering process involving, in
this case, a fermion anti-fermion pair and a photon, so a decay
$1\to 2$ or recombination $2\to 1$. To these external edges we can
assign quantum numbers, specifying the spin, mass, momenta and
other characteristics of the particles involved in the scattering
process.

The set of  edges decomposes in this obvious manner into internal
and external ones $\Gamma^{[1]}=\Gamma^{[1]}_{\rm
int}\cup\Gamma^{[1]}_{\rm ext}$. To calculate the actual
contribution of a graph $\Gamma$, one needs Feynman rules, which
can be heuristically derived from the Lagrangian of the theory in
a straightforward way. They come with a surprise though:
typically, in sensible quantum field theories they do not seem to
make sense, at first sight.

Obviously, we use two different meanings of sense. What goes on
here is that the theories most sensible from a particle physicists
viewpoint are those which agree best with observations. Nature
singles out by this criterion renormalizable quantum field
theories in four dimensions. But then, their Feynman rules seem to
violate common sense: evaluating the Feynman graphs in such
theories by the Feynman rules produces ill-defined quantities
galore. It is a relief then that these senseless quantities
actually make good mathematical sense when one looks at the
structure of graphs quite a bit more closely.

Let us go back to the example of the graph $\Gamma$, regarded as a
QED graph in four dimensions.  Let us describe the structure of
the ill-defined quantities we get from this graph. First of all,
we assign a variable $k_e$ to each edge $e\in\Gamma^{[1]}$.
Variables attached to internal edges we call internal momenta,
while variables attached to external edges we call external
momenta, which we assume to be fixed and given as part of the
quantum numbers of external particles.

Each vertex in the set $\Gamma^{[0]}$ of vertices of $\Gamma$
imposes a constraint on these variables, such that the momenta
attached to a vertex add to zero. One easily recognizes that the
number of free variables left is then equal to the number of loops
in the graph. Those free variables, corresponding to internal
unobserved momenta, have to be integrated out. The Feynman rules
attach propagators $P^{-1}(k_e)$ to each edge $e$, and the edge
variables $k_e$ have to be integrated over  a $D$-dimensional
Euclidean space (as far as short-distance singularities go we can
indeed avoid the complications provided by other signatures of the
metric, or by some non-vanishing curvature). Depending on the
scaling degree $\omega_P$ of the inverse propagators $P(k_e),
P(\lambda k_e)=\lambda^{\omega_P}P(k_e)$ for large $k_e$, this
might or might not be a well-defined integral. This can be easily
decided by powercounting, and leads us to the notion of a degree
of divergence: assigning weights $\omega_P$ to edges (and, in
general, also to vertices), allows, by sole consideration of these
weights and the number of loops in a graph, to decide in advance
if the integrals attached to a graph will have short distance (UV)
divergences. Such an integral is typically of the form
\begin{equation}\phi(\Gamma)=\int \prod_{e\in
\Gamma^{[1]}_{\rm int}} d^D
k_{e}\;P^{-1}(k_{e})\prod_{v\in\Gamma^{[0]}}\delta^K\left(\sum_{j\in
f_v}k_j\right) g(v),\label{int}\end{equation} where $f_v$ is the
set of edges attached to $v$ and $g(v)$ is the factor which the
Feynman rules assign to the vertex $v$, and an appropriate
ordering of the factors along fermionic lines and so on, if
necessary, is understood. Note that this integral representation
implies momentum conservation for the external momenta.

Understanding the singularity structure of such an expression
amounts to an identification of singular subintegrals, which can
possibly be provided only by subgraphs which contain closed loops,
and it thus suffices to consider 1PI graphs and their disjoint
unions to identify all singular subsectors.

So then, what is the message for our example? It turns out that
there is one divergent subgraph for QED  in $D=4$ dimensions. So
what we get is an ill-defined quantity containing another
ill-defined quantity as a subintegral.

How do we get sense into this? There are two steps in this
process, the first is to understand how to make sense out of
graphs which have no divergent subgraphs. The second and harder is
to understand how to do it when subproblems are present. In
between lies the step to understand why divergent subgraphs make
life so much harder.

Consider $$\Gamma_0:=\vc.$$ This is a QED graph which has no
divergent subgraph in four dimensions. By the above such a graph
can be written in the form
\begin{equation}\phi(\Gamma)(m;p_i)=\int_0^{\infty}
\frac{F_\Gamma(r;m;p_i)}{r}dr,\label{asym}\end{equation} with \be
F_\Gamma(0;m;p_i)=0,\;\lim_{r\to\infty}F_\Gamma(r;m;p_i)=Q(\log(r)),\ee
where $Q(\log(r))$ is a polynomial in $\log(r)$ with coefficients
independent of $m;p_i$ and we let $r=\sum_i|k_{e_i}|$, say.

Hence, all what is sick about this graph remains invariant when we
vary these external parameters - the disease is localized, hence
curable: the difference
\begin{equation}\phi(\Gamma)(m;p_i)-\phi(\Gamma)(m;\tilde{p_i})\end{equation}
 exists for any modified external momenta $\tilde{p_i}$.
Actually, in a log divergent graph free of subdivergences the
divergence remains invariant under any diffeomorphism $\psi$ of
external parameters $(m;p_i)\to\psi((m;p_i))$.

So we can give no absolute meaning to the value of a Feynman
graph, but the relative value defined by comparison with another
graph with modified continous quantum numbers exists. A typical
example for a graph without subdivergences is a one-loop graph,
obviously. And that were the early successes of QFT indeed: the
comparison of observables distinguished by different external
parameters.

One point is worth mentioning here: it is not the loop number
which makes a simple subtraction sufficient, but the fact that
there are no subdivergences. That is one of the crucial advantages
of the Hopf algebraic desciption of short distance singularities:
the number of divergent sectors provides a well-defined grading on
that Hopf algebra, and induction over that grading provides a much
clearer understanding how to achieve finite results. With respect
to this grading, the bidegree as we will call it, divergent graphs
free of subdivergences are of bidegree one, and correspond to the
primitive elements in the Hopf algebra. We will often call them
primitive graphs. Ultimately, they are the building blocks out of
which we can assemble the full perturbative expansion, once we
learn how to insert them into each other.

This story has a Lagrangian version: the reference to a chosen
scheme is established by plugging counterterms into the
Lagrangian, such that all Green functions vanish at this reference
'point', from now on called renormalization point. The choice of
this point corresponds to a choice of a subtraction scheme $R$.
Linguistically, we are rather lax: the choice of any scheme like
minimal subtraction, momentum scheme, on-shell scheme and so on
will be allowed. Any such choice, as we will see, corresponds to
the choice of a certain element in the group of characters of the
Hopf algebra, and hence indeed to a point in that group.

What goes wrong when subdivergences are present is obvious -
simple differences like the above will fail. Indeed, the presence
of divergences generates a dependence of the illness on external
parameters. End of theory?

Fortunately not. Let us consider what happens when we take two
primitively divergent graphs and insert them into each other, say
we insert $$\gamma:=\vs$$ into $\Gamma_0$ so that $\Gamma$ is
obtained. Evaluating by the Feynman rules, the integral
$\phi(\gamma)$ will appear as a subintegral of $\phi(\Gamma)$.
Typically, the continous parameters -momenta- attached to the
external legs of $\gamma$ will be integrated over in that larger
integral $\phi(\Gamma)$. But really, what we should insert in that
larger integral is $\phi(\gamma)$ minus its value at the
renormalization point. That is the trick actually: the elimination
of subdivergences goes first, before the cure is available for the
larger problem posed by the larger graph $\Gamma$. This is
consistent with the Lagrangian story: curing the sickness of
$\gamma$ required the insertion of its counterterm into the
Lagrangian. Thus this modified Lagrangian will, whenever providing
$\gamma$, also provide its counterterm, hence provide the cured
version of $\gamma$.

Summarizing, $\Gamma$ in our example contains one interesting
subgraph, the one-loop self-energy graph $\gamma$. It is the only
subgraph which provides a divergence, and the whole UV-singular
structure comes from this subdivergence and from the overall
divergence of $\Gamma$ itself.  From the analytic expressions
corresponding to $\Gamma$, to $ \Gamma_0$ and to $\gamma$ we can
form the analytic expression corresponding to the renormalization
of the graph $\Gamma$. It is given by \be
\phi(\Gamma)-R(\phi(\Gamma))-R(\phi(\gamma))\phi(\Gamma_0)+R\left(R(\phi(\gamma))\phi(\Gamma_0)
\right).\ee We emphasize that the crucial step in obtaining this
expression is the use of the graph $\Gamma$ and its disentangled
pieces, $\gamma$ and $\Gamma_0=\Gamma/\gamma$. Diagrammatically,
the above expression reads (omitting $\phi$) {\large
$$ \vv-R\left(\vv\right)-R\left(\vs\right)\vc$$ $$+R\left(R\left(\vs\right)\vc\right).$$}

The unavoidable arbitrariness in the so-obtained expression lies
in the choice of the map $R$ which we suppose to be such that it
does not modify the short-distance singularities (UV divergences)
in the analytic expressions corresponding to the graphs. It just
evaluates graphs at the chosen renormalization point, so it
employs the chosen scheme. Certain requirements on $R$ have to be
demanded \cite{Chen,RHI,review}: it has to be faithful to
short-distance singularities, and it has to establish a Baxter
algebra on the target space of the Feynman rules $\phi:H\to V$:
\be R(ab)+R(a)R(b)=R(aR(b))+R(R(a)b),\;R:V\to V,\; a,b\in
V.\label{Baxter}\ee This then renders the above combination of
four terms finite. If there were no subgraphs, a simple
subtraction $\phi(\Gamma)-R(\phi(\Gamma))$ would suffice to
eliminate the short-distance singularities, but the necessity to
obtain local counterterms forces us to first subtract
subdivergences. This is Bogoliubov's famous $\bar{R}$ operation
\cite{Collins}, which delivers here: \be\phi(\Gamma)\to
\bar{R}(\phi(\Gamma))=\phi(\Gamma)-R(\phi(\gamma))\phi(\Gamma_0).\ee
This provides two of the four terms above. Amongst them, these two
are free of subdivergences and hence provide only a local overall
divergence. The projection of these two terms into the range of
$R$ provides the other two terms, which combine to the counterterm
\be Z_\Gamma=-R(\phi(\Gamma))+R(R(\phi(\gamma))\phi(\Gamma_0))\ee
of $\Gamma$, and addition of this counterterm delivers the finite
result above, in the kernel of $R$, by the fact that the UV
divergences are not changed by the renormalization map $R$.

Locality is indeed connected to the absence of subdivergences: if
a graph has a sole overall divergence, UV singularities only
appear when all loop momenta tend to infinity jointly. Regarding
the analytic expressions corresponding to a graph as a Taylor
series in external parameters like masses or momenta,
powercounting establishes that only the coefficients of the first
few polynomials in these parameters are UV singular. Hence they
can be subtracted by a counterterm which is a polynomial in fields
and their derivatives. The argument fails as long as one has not
eliminated all subdivergences: their presence can force each term
in the Taylor series to be divergent. For example, if none of the
edges or vertices of the subgraph involves the external momenta
(by routing external momenta so that they avoid the subgraph under
consideration), then no derivative with respect to those
parameters can possibly eliminate the divergence generated by this
subgraph. Hence, the preparation of a graph for a local
subtraction by Bogoliubov's operation is unavoidable. The
independence of the singularities of a prepared graph on the
variation (diffeomorphism) of external parameters is a strong hint
to regard the remaining singularity as a residue, an analogy with
far-reaching consequences \cite{RHII} to which we will come back
below.

The basic operation so far was the disentanglement of the graph
$\Gamma$ into pieces $\gamma$ and cographs
$\Gamma_0=\Gamma/\gamma$, and this very disentanglement gives rise
to a Hopf algebra structure, as was first observed in \cite{DK1},
which we will describe shortly.

It is useful to study the invariants of a permutation of places
where a subgraph $\gamma$ is inserted in a graph $\Gamma_0$. What
obviously remains invariant is the hierarchical structure of
subdivergences, what varies is the topology of the graph. Indeed,
the counterterm for $\gamma$ provided by the Lagrangian is the
same wherever we insert $\gamma$, and the difference of two such
insertions will need no counterterm for $\gamma$. This has
immediate consequences for number-theory \cite{BKK} to be
commented  at the end of section seven, when we connect such
invariance to Galois symmetry.

For now, as an example, consider the two graphs {\large
\be\Gamma_1=\vv,\;\Gamma_2=\vvm.\label{intro}\ee} They have one
common property: both of them can be regarded as the graph {\large
\be\Gamma_0=\Gamma_1/\gamma=\Gamma_2/\gamma=\vc,\ee} into which
the subgraph {\large $$\gamma=\vs$$} is inserted, at two different
places $i,j$, though.  Such graphs are equivalent, in the sense
that the combinatorial process of renormalization produces exactly
the same subtraction terms for both of them \cite{DK1}. This
equivalence can be most meaningful stated using the language of
operads \cite{L}: inserting a subgraph at different places is an
operad composition, with a labelled composition to denote the
places where to insert subgraphs. Vanishing of the leading
singularity for that difference then means that the permutation
group for that operad composition  is trivially represented on
that leading short-distance singularity.

The combinatorics of renormalization is essentially governed by
this bookkeeping process of the hierarchies of subdivergences, and
this bookkeeping is what is delivered by rooted trees. They are
just the appropriate tool to store the hierarchy of disjoint and
nested subdivergences, and ultimately, overlapping subdivergences,
which resolve into the former \cite{DK1,overl,Chen}.

Hence the Hopf algebra of Feynman graphs indeed has a role model:
the Hopf algebra of rooted trees. Rooted trees can be assigned in
two natural ways to a Feynman graphs:\\ i) decomposing momentum
space Feynman integrals into divergent sectors \cite{overl}. This
amounts to a resolution of overlapping divergences into disjoint
or nested sectors in the integral representation of graphs
provided by the momentum space Feynman rules. This can be done,
and will be exhibited later on when we comment on how to use
Hochschild
cohomology to provide a proof for renormalizability.\\
ii) on the other hand, starting from coordinate space Feynman
rules, the singularities stratify in rooted trees directly, upon
the fact that they are supported along diagonals in the
configuration space of the location of vertices.

Combining both viewpoints, the Fourier transform becomes a map
between two rooted tree Hopf algebras. For now, we restrict
ourselves to shortly describe the configuration space viewpoint on
Feynman graphs, before we finally define the Hopf algebra of
Feynman graphs more formally.

First, let us ask what to make out of graphs which have
overlapping divergences.  For the non-overlapping graphs
$\Gamma_1,\Gamma_2$ above there is a unique way to obtain them
from {\large
$$\Gamma_0=\Gamma_1/\gamma=\Gamma_2/\gamma=\vc$$} and the
one-loop vertex correction $\gamma$. We plug $\gamma$ into the
other vertex-correction at an appropriate internal vertex to
obtain the desired graphs.  On the other hand, for a graph which
contains overlapping divergences we have typically no unique
manner, but several ways instead, how to obtain it. It is in this
loss of uniqueness how to decompose it, or how to assemble it from
its parts, where some of the most fascinating aspects of QFT
reside: indeed, we will sketch at the end of this paper that we
are fighting here with the famous problem of unique factorization,
and will argue that the cure is quite similar to what one does for
algebraic number fields: factorize into prime ideals.

Consider Fig.(\ref{f1}). It shows two ways of obtaining a sum
$\Omega$ of two graphs, by inserting a vertex graph $\Gamma_1$
into the two internal vertices $a,b$ of a self-energy $\omega_1$,
or a vertex graph $\Gamma_2$ into $\omega_2$, \be
\Omega=\omega_1\star\Gamma_1=\omega_2\star\Gamma_2.\ee Note that
each of the two graphs $\Omega_1,\Omega_2$ in $\Omega$ has four
internal vertices. There are two subsets of three vertices in each
which belong to the two divergent vertex subgraphs we can identify
in them. In coordinate space, these subsets provide a singular
strata located along the corresponding diagonals where these
subsets collapse to a single point. The coordinate space Feynman
rules do not make sense along these diagonals, and their
continuation to these diagonals is a problem dual to the
compactification of configuration space along diagonals. The
choice of compactification corresponds to the choice of a
renormalization scheme, and the fact that diagonals in
configuration space come stratified by rooted trees \cite{FMcPh}
invites to establish the Hopf algebra techniques of
renormalization theory in the study of configuration spaces, and
vice versa. \bookfig{f1.eps}{f1}{f1}{This sum $\Omega$ of two
graphs $\Omega_1,\Omega_2$ can be obtained in two different ways.
We can either glue $\Gamma_1$ into $\omega_1$ by identifying the
external edges of $\Gamma_1$ with the edges adjacent to the
vertices $a$ or $b$ of $\omega_1$, or similarly, do so for the
graphs $\omega_2,\Gamma_2$, with the same result $\Omega$.}{6} The
short distance singularities of Feynman graphs then come solely
from  regions where all vertices are located at coinciding points.
One has no problem to define the Feynman integrand in the
configuration space of vertices at distinct locations, while a
proper extension to diagonals is what is required \cite{EG,S}.

Due to the Hopf algebra structure of Feynman graphs we can define
the renormalization of all such sectors without making recourse to
any specific analytic properties of the expressions (Feynman
integrals) representing those sectors. The only assumption we make
is that in a sufficiently small neighborhood of such an ultralocal
region (the neighborhood of a diagonal) we can define the scaling
degree, --the powercounting--, in a sensible manner: combinatorics
suffices. Analytic detail arising from quests for causality for
example can be imposed later. This principle indeed goes far:
almost all detail about the Feynman rules for propagators and
vertices which goes beyond their scaling degree is unnecessary at
this stage. Having determined these powercounting degrees and
having chosen a renormalization scheme, we will soon formally
define the principle of multiplicative subtraction which will tell
us how we get local counterterms and finite renormalized
quantities whatever the finer detail is of the Feynman rules. This
is the strong combinatorial backbone of QFT, which ultimately has
its source in the notion of a residue, and in the fundamental
invariance properties of residues.

This is indeed a very useful application of the Hopf algebra: the
decomposition into its primitives reveals the range over which the
combinatorial structures of renormalization are stable to include,
for example, fluctuations of the metric as long as the scaling
degree of propagators is microlocally unchanged, as such
fluctuation do not alter the residue of a bidegree one graph. This
allows to understand the recent results of Brunetti and
Fredenhagen \cite{BF} from a combinatorial viewpoint.

Its time now to define the Hopf algebra of Feynman graphs, which
expresses this combinatorial backbone of renormalization theory.

\subsection{Lie and Hopf algebras of Feynman graphs}
We start giving some formal definitions. Our presentation follows
\cite{CKnew}.

 First, we define $n$-particle irreducible ($n$-PI) graphs.
 We will exclude graphs with self-loops: no edge connects a vertex
 to itself. But we will allow for two different vertices to be connected by several
 edges. Note that self-loops are naturally excluded in a massless
 theory.
\begin{defn}
A $n$-particle irreducible graph ($n$-PI graph) $\Gamma$ consists
of edges and vertices, without self-loops, such that upon removal
of any set of $n$ of its edges it is still connected. Its set of
edges is denoted by $\Gamma^{[1]}$ and its set of vertices is
denoted by $\Gamma^{[0]}$. Edges and vertices can be of various
different types.
\end{defn}
The type of an edge is often indicate by the way we draw it:
(un-)oriented straight lines, curly lines, dashed lines and so on.
These types of edges, often called propagators in physicists
parlance, are chosen with reference to Lorentz covariant wave
equations: the propagator as the analytic expression assigned to
an edge is an inverse wave operator with boundary conditions
typically chosen in accordance with causality. We can, if desired,
ignore such consideration by the choice of an Euclidean metric.

The types of vertices are determined by the types of edges to
which they are attached:
\begin{defn}
For any vertex $v$ $\in \Gamma^{[0]}$ we call the set
$f_v:=\{f\in\Gamma^{[1]}\mid v\cap f\not=\emptyset \} $ its type.
\end{defn}
Note that $f_v$ is a set of edges.

 Of particular importance are
the 1PI graphs. They do not decompose into disjoint graphs upon
removal of an edge. Note that any $n$-PI graphs is also
$(n-1)$-PI, $ \forall n\geq 2$. A graph which is not 1-PI is
called reducible. Also, any connected graph is considered as 0-PI.

 A further notion needed
is the one of external and internal edges, introduced before.
\begin{defn}
An edge $f$ $\in \Gamma^{[1]}$ is internal, if $\{v_f\}:=f\cap
\Gamma^{[0]}$ is a set of two elements.
\end{defn}
So, internal edges connect two vertices of the graph $\Gamma$.
\begin{defn}
An edge $f$ $\in \Gamma^{[1]}$ is external, if $f\cap
\Gamma^{[0]}$ is a set of one element.
\end{defn}
As we exclude self-loops, this means that an external edge has an
open end. Thus external edges are associated with a single vertex
of the graph. These edges correspond to external particles
interacting in the way prescribed by the graph.

We now turn to the possibilities of inserting graphs into each
other. Our first requirement is to establish bijections between
sets of edges so that we can define gluing operations.
\begin{defn}
We call two sets of edges $I_1,I_2$ compatible, $I_1\sim I_2$, iff
they contain the same number of edges, of the same type.
\end{defn}
\begin{defn}
Two vertices $v_1,v_2$ are of the same type, if $f_{v_1}$ is
compatible with $f_{v_2}$.
\end{defn}
Quite often, we will shrink a graph to a point. The only useful
information still available after that process is about its set of
external edges:
\begin{defn}
We define ${\bf res}(\Gamma)$ to be the result of identifying
$\Gamma^{[0]}\cup\Gamma^{[1]}_{\rm int}$ with a point in $\Gamma$.
\end{defn}
An example is $$\res.$$ Note that ${\bf res}(\Gamma)^{[1]}\equiv
{\bf res}(\Gamma)^{[1]}_{\rm ext}\sim \Gamma^{[1]}_{\rm ext}$. By
construction all graphs which have compatibe sets of external
edges have the same residue.

If the set $\Gamma^{[1]}_{\rm ext}$ is empty, we call $\Gamma$ a
vacuumgraph, if it contains a single element we call the graph a
tadpole graph. Vacuum graphs and tadpole graphs will be discarded
in what follows. If $\Gamma^{[1]}_{\rm ext}$ contains two
elements, we call $\Gamma$ a self-energy graph, if it contains
more than two elements, we call it an interaction or vertex graph.
Further we restrict ourselves to graphs which have vertices such
that the cardinality of their types is $\geq 2$. In the presence
of external fields, this can be relaxed.

At this stage, we are  provided with a list of edges and vertices
obtained typically, but not necessarily, from some QFT Lagrangian.
Weights assigned to these elements allow to discriminate graphs
formed from these objects by the means of powercounting. We can
thus meaningfully speak about 1PI graphs which are superficially
divergent or convergent - graphs $\Gamma$ such that the integral
Eq.(\ref{int}) attached to $\Gamma$ diverges or converges at the
upper boundary of the final $r=\sum_{e\in \Gamma^{[1]}_{\rm
int}}|k_e|$ integration in Eq.(\ref{asym}).

We are after a mechanism which eliminates all possible divergences
coming from subintegrations. These can be detected by
powercounting on the 1PI subgraphs, and we will soon introduce a
Hopf algebra structure based on superficially divergent 1PI
graphs.

\subsection{External structures}
Let us mention here one more useful notational device: the
external structures of \cite{RHI}. To each external edge
$f\in\Gamma^{[1]}_{\rm ext}$ we assign an external momentum $p_f$,
a vector in some (to be appropriately specified) vectorspace and
impose the condition $\sum_{f\in\Gamma_{\rm ext}^{[1]}}p_f=0$. We
let $E_{{\bf res}(\Gamma)}$ be the linear space of functions of
those variables $p_f$ fulfilling this condition.

Following \cite{RHI} we use a notation familiar from distribution
theory to describe the evaluation of a graph $\Gamma$ by some
Feynman rules $\phi:{\cal H}\to E_{{\bf res}(\Gamma)}$: with
$\phi(\Gamma)\in E_{{\bf res}(\Gamma)}$, we denote by
$\langle\sigma,\phi(\Gamma)\rangle$ the evaluation of
$\phi(\Gamma)$ with respect to the distribution $\sigma$ on
$E_{{\bf res}(\Gamma)}$, in the same way as $\langle
\delta_a,f\rangle=f(a)$ defines the evaluation of a function at
$a$ by the pairing with a Dirac $\delta$-distribution supported at
$a$.

A graph $\Gamma$ together with a distribution $\sigma$ we call a
specified graph, and write $(\Gamma,\sigma)$ for such a pair. We
can regard it as a graph with a further prescription how to
evaluate it.

We will need these notions to keep track of our perturbative
expansion: let us assume that the Lagrangian $L$ which is the
source for our perturbative expansion is a sum over $k$ field
monomials. Assume that $L$ contains $n$ different fields
$\phi_1,\ldots\phi_n$. Then to each monomial $P(\{\phi_i\})$ in
$L$ we can assign the sequence $(i_1,\ldots,i_n)_P$, which tells
us the degree of $P$ in each field. We call this sequence the
field degree of $P$. If two monomials $P,P^\prime$ deliver equal
field degrees, they both can be described by graphs with an equal
residue in the above sense: their corresponding Feynman graphs
$\Gamma$ will have identical external legs, hence identical
residues ${\bf res}(\Gamma)$. Counterterms for such monomials are
then calculated by suitable projections in $E_{{\bf
res}(\Gamma)}$, implemented by distributions
$\sigma_P,\sigma_{P^\prime}$.

A typical example are the mass and wave function renormalization
of a scalar propagator, with monomials $P=-m^2\phi^2/2$ and
$P^\prime=\partial_\mu\phi\partial^\mu\phi/2$, say. Both monomials
are quadratic in a scalar field $\phi$ and the contributions of
two-point graphs to $P,P^\prime$ are obtained from
$$ \langle \sigma_P,\phi(\Gamma)\rangle$$ and
$$ \langle \sigma_{P^\prime},\phi(\Gamma)\rangle,$$
with $\langle\sigma_P,f(m^2,q^2)\rangle=\partial_{m^2}
f\mid_{m^2=q^2}$ and
$\langle\sigma_{P^\prime},f(m^2,q^2)\rangle=\partial_{q^2}
f\mid_{m^2=q^2}$, for any function $f(m^2,q^2)$. Similarly, one
extends to other cases for monomials of equal field degree, and
one can conveniently express the formfactor decomposition of Green
functions in this language.

In such a notation, all field monomials $P$ in the Lagrangian
correspond to expressions $\langle \sigma_P,\phi({\bf
res}(\Gamma))\rangle$, where $\Gamma$ is a graph which external
legs in accordance with the field degree of $P$, and one always
has that \be\langle \sigma_P, \phi(\Gamma)\rangle =\rho_P(\Gamma)
\langle \sigma_P,\phi({\bf res}(\Gamma))\rangle,\ee where
$\rho_P(\Gamma)$ is a scalar function (the form-factor) which
fulfills that on disjoint diagrams $\rho_P(\Gamma_1\Gamma_2)\equiv
\rho_P(\Gamma_2\Gamma_1)=\rho_P(\Gamma_1)\rho_P(\Gamma_2)$,
$\forall P$, regardless of the fact that $\langle \sigma_P,
\phi(\Gamma_P)\rangle$ can be matrix-valued. The simple fact that
the action is a Lorentz scalar indeed mandates that a formfactor
decomposition into scalar coefficients is always possible, and is
in accordance with the product structure of graphs: the evaluation
of a product is the commutative product of the evaluations.

It is those scalar coefficients which appear as characters on the
Hopf algebra of graphs which we describe below. Our route will be
to first define a pre-Lie product on graphs, which at the same
time establishes a Lie bracket upon its antisymmetrization, and
then take the dual of the  universal enveloping algebra of the
so-obtained Lie algebra as the Hopf algebra under consideration.
This indeed provides the always commutative  Hopf algebra which
underlies the disentanglement of graphs into their divergent
parts, the combinatorial backbone of local quantum field theory.

\subsection{The Pre-Lie Structure}
Consider two graphs $\Gamma_1,\Gamma_2$. Assume that $\Gamma_2$ is
an interaction graph. For a chosen vertex $v_i\in\Gamma_1^{[0]}$
such that $f_{v_i}\sim \Gamma^{[1]}_{2,\rm ext}$, we define
\be\Gamma_1\star_{v_i}\Gamma_2=\Gamma_1/v_i\cup\Gamma_2/\Gamma^{[1]}_{2,\rm
ext},\ee where in the union of these two sets we create a new
graph such that for every edge $f_j\in f_{v_i}$, $\{v_{f_j}\}$
contains precisely one element in $\Gamma_2^{[0]}$. Then we sum
over all possible bijection between $f_{vi}$ and $\Gamma_{2,\rm
ext}^{1}$, and normalize such that topologically different graphs
are generated precisely once.

We now define
\be\Gamma_1\star\Gamma_2=\sum_{w\in\Gamma_1^{[0]}\atop f_w\sim
\Gamma^{[1]}_{2,\rm ext}} \Gamma_1\star_w\Gamma_2.\ee All this can
be easily generalized to the insertion of self-energy graphs,
replacing internal edges by self-energy graphs which have the
corresponding external edges.

We also define the insertion of specified graphs in the similar
vein, by requiring that
\be(\Gamma_1,\sigma_1)\star(\Gamma_2,\sigma_2)=(\Gamma_1\star
\Gamma_2,\sigma_1).\ee That is, by inserting a graph, we forget
about conditions imposed on its external legs. Dually, for the
Hopf algebra below, this corresponds to the fact that upon
disentangling a graph, we have the freedom to impose constraints
-renormalization conditions- to its subgraphs.
\begin{prop}
The operation $\star$ is pre-Lie:
\begin{eqnarray*}[(\Gamma_1,\sigma_1)\star(\Gamma_2,\sigma_2)]\star(\Gamma_3,\sigma_3) & - &
(\Gamma_1,\sigma_1)\star[(\Gamma_2,\sigma_2)\star(\Gamma_3,\sigma_3)]\\
=[(\Gamma_1,\sigma_1)\star(\Gamma_3,\sigma_3)]\star(\Gamma_2,\sigma_2)
& - &
(\Gamma_1,\sigma_1)\star[(\Gamma_3,\sigma_3)\star(\Gamma_2,\sigma_2)]
.\end{eqnarray*}
\end{prop}
To sketch the proof, which is elementary for unspecified graphs,
we note that the insertion of subgraphs is a local operation, and
that on both sides, the difference amounts to plugging the
subgraphs $\Gamma_2,\Gamma_3$ into $\Gamma_1$ at disjoint places,
which is evidently symmetric under the exchange of $\Gamma_2$ with
$\Gamma_3$. For specified graphs, we choose a color for each
evaluation $\sigma_i$ under consideration, and represent the pair
$(\sigma,\Gamma)$ by a colored graph. Coloring does not spoil the
locality of the insertions.

If needed, this can be generalized in a way as to maintain
constraints on external legs when inserting a graph into another,
still maintaining the pre-Lie structure. Integration over the
edges according to Eq.(\ref{int}) is then done in a way such that
it obeys the constraints valid for each subgraph. Such extensions
are useful in practice when one desires to decompose graphs into
various analytic subfactors \cite{Book}, and underly the
decomposition into primitives discussed in section five.

\noindent In most of what follows we only use unspecified graphs,
with the generalization to specified graphs being evident.

We get a Lie algebra ${\cal L}$ by antisymmetrizing this
operation, \be
[\Gamma_1,\Gamma_2]=\Gamma_1\star\Gamma_2-\Gamma_2\star\Gamma_1\label{Lie}\ee
and a Hopf algebra ${\cal H}$ as the dual of the universal
enveloping algebra of this Lie algebra. Typically, one restricts
attention to graphs which are superficially divergent, with
residues corresponding to field monomials in the Lagrangian.

Superficially convergent graphs  can be incorporated using the
trivial abelian Lie algebra which they span when one regards them
as specified graphs. The fact that they do not contribute to
counterterms in the Lagrangian means that they are annihilated by
external structures which project onto the contributions to
superficially divergent field monomials $P$. Hence they have a
vanishing Lie bracket among themselves and furthermore form a
semi-direct product with their superficially divergent cousins
\cite{RHI}. An example of the Lie bracket Eq.(\ref{Lie}) is
provided by Fig.(\ref{f2}). \bookfig{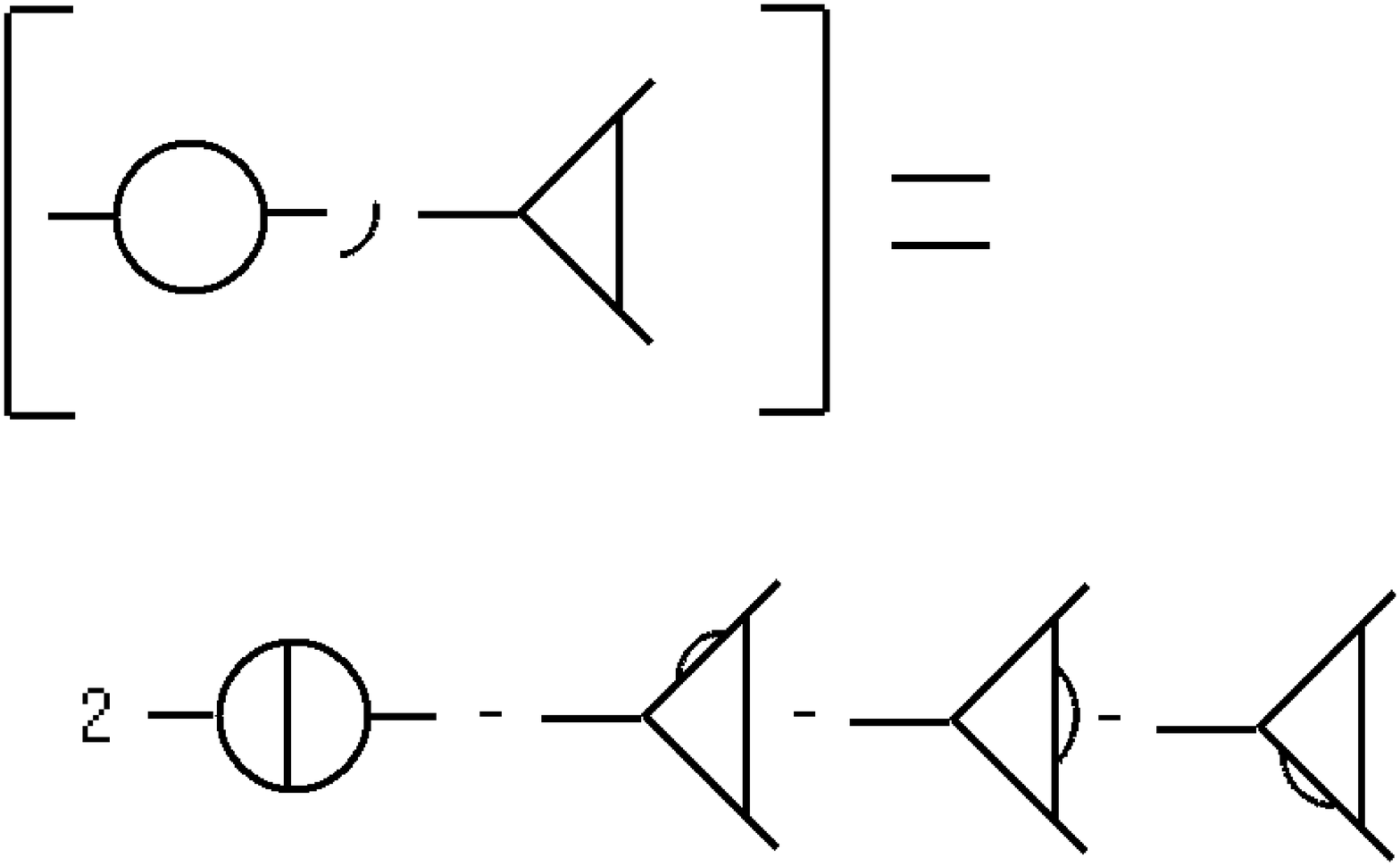}{f2}{f2}{The
Lie-bracket
$[\Gamma_1,\Gamma_2]=\Gamma_1\star\Gamma_2-\Gamma_2\star\Gamma_1$.}{6}

\subsection{The principle of multiplicative subtraction}
Having defined Lie algebra structures on graphs, it is now easy to
harvest them to give a clear conceptual meaning to the
renormalization process. As announced, we just have to dualize the
universal enveloping algebra ${\cal U(L)}$ of ${\cal L}$ and
obtain a commutative, but not cocommutative Hopf algebra ${\cal
H}$ \cite{RHI}.

From now on, when we want to distinguish carefully between the
Hopf and Lie algebras of Feynman graphs we write $\delta_\Gamma$
for the multiplicative generators of the Hopf algebra and write
$Z_\Gamma$ for the dual basis of the universal enveloping algebra
of the Lie algebra ${\cal L}$ with pairing \be \langle
Z_\Gamma,\delta_{\Gamma^\prime}\rangle=\delta^K_{\Gamma,\Gamma^\prime},\ee
where on the rhs we have the Kronecker $\delta^K$, and extend the
pairing by means of the coproduct \be\langle
Z_{\Gamma_1}Z_{\Gamma_2},X\rangle =\langle Z_{\Gamma_1}\otimes
Z_{\Gamma_2},\Delta(X) \rangle.\ee

Quite often, we want to refer to the graph(s) which index an
element in  ${\cal H}$ or ${\cal L}$. For that purpose, for each
element in ${\cal H}$ and each element in ${\cal L}$ we introduce
a map to graphs: \be \overline{Z_{X}}= X,\overline{ \delta_{X}}=
X.\ee

As we already have emphasized the Hopf algebra of rooted trees is
the role model for the Hopf algebras of Feynman graphs which
underly the process of renormalization when formulated
perturbatively at the level of Feynman graphs. The following
formulas should be of no surprise for the reader acquainted with
the universal Hopf algebra of rooted trees and are a
straightforward generalization of similar formulas for rooted tree
Hopf algebras \cite{CK1}.

First of all, we start considering one-particle irreducible graphs
as the linear generators of the Hopf algebra, with their disjoint
union as product. We then identify  the  Hopf algebra as described
above  by a coproduct $\Delta:{\cal H}\to {\cal H}\otimes{\cal
H}$: \be\Delta(\Gamma)=\Gamma\otimes
1+1\otimes\Gamma+\sum_{\gamma{\subset}
\Gamma}\gamma\otimes\Gamma/\gamma,\ee where the sum is over all
unions of one-particle irreducible (1PI) superficially divergent
proper subgraphs and we extend this definition to products of
graphs so that we get a bialgebra. The above sum should, when
needed, also run over appropriate external structures to specify
the appropriate type of local insertion \cite{RHI} which appear in
local counterterms, which we omitted in the above sum for
simplicity. Fig.(\ref{f3}) gives an example of a coproduct.
\bookfig{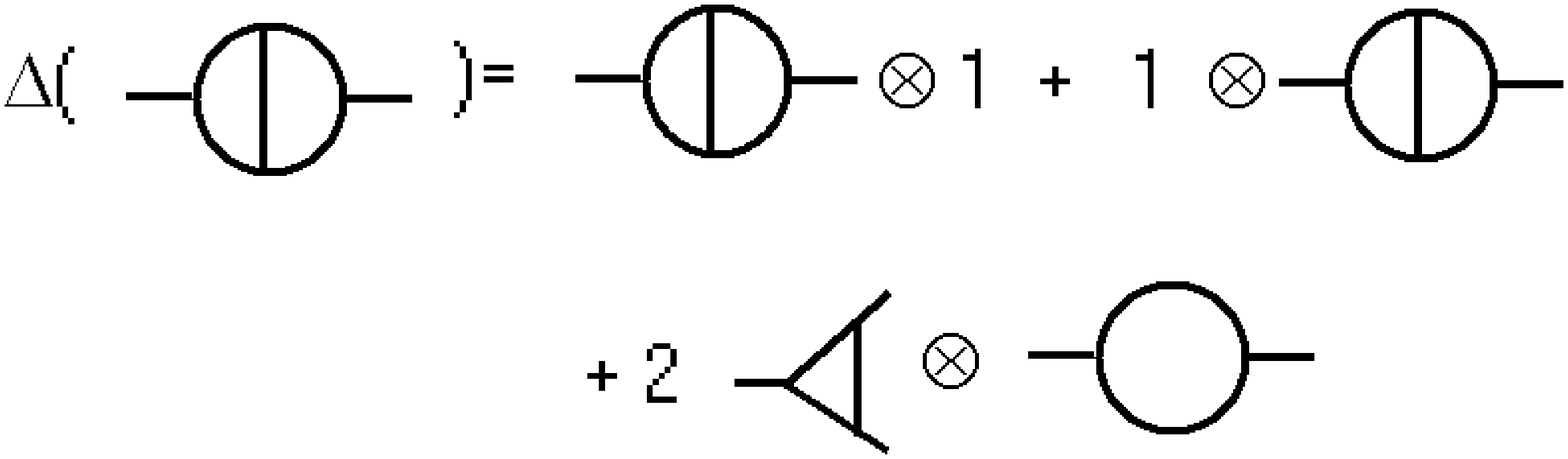}{f3}{f3}{The coproduct $\Delta(\Gamma)$.}{3.9}

For any Hopf algebra element $X$ we often write a shorthand for
its coproduct $$ \Delta(X)=\widetilde{\Delta}(X)+X\otimes
1+1\otimes X=X\otimes 1+1\otimes X+X^\prime\otimes
X^{\prime\prime}.$$ Let now $X$ be a 1PI graph. For each term in
the sum $\widetilde{\Delta}(X)=\sum_i X^\prime\otimes
X^{\prime\prime}$ we have unique gluing data $G_i$ such that \be
X=X^{\prime\prime}\star_{G_i}X^\prime,\;\forall i.\label{glue}\ee
These gluing date describe the necessary bijections to glue the
components $X^\prime$ back into $X^{\prime\prime}$ so as to obtain
$X$.

The counit $\bar{e}$ vanishes on any non-trivial Hopf algebra
element, $\bar{e}(1)=1,\,\bar{e}(X)=0$. At this stage we have a
commutative, but typically not cocommutative bialgebra. It
actually is a Hopf algebra as the antipode in such circumstances
comes for free as \be S(\Gamma)=-\Gamma-\sum_{\gamma{\subset}
\Gamma}S(\gamma)\Gamma/\gamma.\ee

The next thing we need are Feynman rules, which we regard as maps
$\phi:{\cal H}\to V$ from the Hopf algebra of graphs ${\cal H}$
into an appropriate space $V$.

Over the years, physicists have invented many calculational
schemes in perturbative quantum field theory, and hence it is of
no surprise that there are many choices for this space.

For example, if we want to work on the level of Feynman integrands
in a BPHZ scheme, we could take as this space a suitable space of
Feynman integrands (realized either in momentum space or
configuration space, whatever suits). An alternative scheme would
be the study of regularized Feynman integrals, for example the use
of dimensional regularization would assign to each graph a
Laurent-series with poles of finite order in a variable $\ve$ near
$\ve=0$, and we would obtain characters evaluating in this ring,
an approach leading to the Riemann-Hilbert decomposition described
below. In any case, we will have
$\phi(\Gamma_1\Gamma_2)\equiv\phi(\Gamma_2\Gamma_1)=\phi(\Gamma_1)\phi(\Gamma_2)$,
$\forall \phi:{\cal H}\to V$.

Then, with the Feynman rules providing a canonical character
$\phi$, we will have to make one further choice: a renormalization
scheme. This is is a map $R:V\to V$, and we demand that is does
not modify the UV-singular structure: in BPHZ language, it should
not modify the Taylor expansion of the integrand for the first
couple of terms divergent by powercounting. In dimensional
regularization, then, we demand that it does not modify the pole
terms in $\ve$. Furthermore, we require it to make the pair
$(V,R)$ into a Baxter algebra, as in Eq.(\ref{Baxter}).

Finally, the principle of multiplicative subtraction emerges: we
define a further character $S_R^\phi$ which deforms $\phi\circ S$
slightly and delivers the counterterm for $\Gamma$ in the
renormalization scheme $R$: \be
S_R^\phi(\Gamma)=-R[\phi(\Gamma)]-R\left[\sum_{\gamma{\subset}
\Gamma}S_R^\phi(\gamma)\phi(\Gamma/\gamma)\right]\ee which should
be compared with the undeformed \be \phi\circ
S=-\phi(\Gamma)-\sum_{\gamma{\subset} \Gamma}\phi\circ S
(\gamma)\phi(\Gamma/\gamma).\ee Then, the classical results of
renormalization theory follow immediately \cite{DK1,overl,CK1}. We
obtain the renormalization of $\Gamma$ by the application of a
renormalized character $$\Gamma\to S_R^\phi\star\phi(\Gamma)$$ and
the $\bar{R}$ operation as \be\bar{R}(\Gamma)=\phi(\Gamma)+
\sum_{\gamma{\subset}\Gamma}
S_R^\phi(\gamma)\phi(\Gamma/\gamma),\ee so that we have \be
S_R^\phi\star\phi(\Gamma)=\bar{R}(\Gamma)+S_R^\phi(\Gamma).\ee

In the above, we have given all formulas in their recursive form.
Zimmermann's original forest formula solving this recursion is
obtained when we trace our considerations back to the fact that
the coproduct of rooted trees can be written in non-recursive
form, and similarly the antipode \cite{overl}.  We will come back
below to this transition between graphs and trees. Also, we note
that the principle of multiplicative subtraction can be formulated
in much larger generality, as it is a basic combinatorial
principle, see for example \cite{Fot} for another appearance of
this principle.

\subsection{The Bidegree} A fundamental notion is the bidegree of
a 1PI graph (see \cite{Foissy} for a convenient review of notions
needed here). Usually, induction in perturbative QFT, aiming to
prove a desired result is carried out using induction over the
loop number, an obvious grading for 1PI graphs. On quite general
grounds, for our Hopf algebras there exists another grading, which
is actually much more useful. We call it the bidegree, ${\rm
bid}(\Gamma)$. To motivate it, consider a superficially divergent
$n$-loop graph $\Gamma$ which has no divergent subgraph. It is
evident that its short-distance singularities can be treated by a
single subtraction, for any $n$. It is not the loop number, but
the number of divergent subgraphs which is the most crucial notion
here. Fortunately, this notion has a precise meaning in the Hopf
algebra of superficially divergent graphs using the projection
into the augmentation ideal. This indeed counts the degree in
renormalization parts of a graph: an overall superficially
convergent graph has bidegree zero by definition, a primitive Hopf
algebra element has bidegree one, and so on.

So we have ${\cal H}=\bigoplus_{i=0}^\infty{\cal H}^{(i)}$, with
${\rm bid}({\cal H}^{(i)})=i$.   To define this decomposition, let
${\cal H}_{\rm Aug}$ be the augmentation ideal of the Hopf
algebra, and let $P: {\cal H}\to {\cal H}_{\rm Aug}$ be the
corresponding projection $P={\rm id}-E\circ \bar{e}$, with
$E(q)=qe$. Let $\widetilde{\Delta}(X)=\Delta(X)-e\otimes
X-X\otimes e$, as before. $\widetilde{\Delta}$ is still
coassociative, and for any $X\in {\cal H}_{\rm Aug}$ there exists
a unique maximal $k$ such that $ \widetilde{\Delta}^{k-1}(X)\in
[{\cal H}^{(1)}]^{\otimes k}.$ Here, ${\cal H}^{(1)}$ is the
linear span of primitive elements $z$: $\Delta(z)=z\otimes
e+e\otimes z$.
\begin{defn}
The map $X\to k$ obtained as described above is called the
bidegree $deg_p$ of $X$.
\end{defn}
The bidegree is conserved under the coproduct and under the
product (disjoint union). Typically, all properties connected to
questions of renormalization theory can be proven more efficiently
using the grading by the bidegree instead of the loop number. In
particular, we will discuss below the interplay between Hochschild
cohomology and the bidegree, which explains locality of
counter\-terms and finiteness of renormalized graphs rather
succinctly. But let us first come to an even more succinct
formulation of renormalization theory, making use of the existence
of complex regularizations.

\section{Renormalization and the Riemann--Hilbert\\ problem}
\subsection{The Birkhoff decomposition}
The Feynman rules in dimensional or analytic regularization
determine a character $\phi$ on the Hopf algebra which evaluates
as a Laurent series in a complex regularization parameter $\ve$,
with poles of finite order, this order being bounded by the
bidegree of the Hopf algebra element to which $\phi$ is applied.
In minimal subtraction, $\phi_-:=S^\phi_{R=MS}$ has similar
properties: it is a character on the Hopf algebra which evaluates
as a Laurent series in a complex regularization parameter $\ve$,
with poles of finite order, this order being bounded by the
bidegree of the Hopf algebra element to which $S^\phi_{R=MS}$ is
applied, only that there will be no powers of $\ve$ which are
$\geq 0$. Then, $\phi_+:=S^\phi_{R=MS}\star\phi$ is a character
which evaluates in a Taylor series in $\ve$, all poles are
eliminated. We have the Birkhoff decomposition
\be\phi=\phi_-^{-1}\star\phi_+.\ee

This establishes  an amazing connection between the
Riemann--Hilbert problem and renormalization \cite{RHI,RHII}. It
uses in a crucial manner once more that the multiplicativity
constraints Eq.(\ref{Baxter}),
$$ R[xy]+R[x]R[y]=R[R[x]y]+R[xR[y]],$$ ensure that the
corresponding counterterm map $S_R$ is a character as well, \be
S_R[xy]=S_R[x]S_R[y],\;\forall x,y\in H,\ee by making the target
space of the Feynman rules into a Baxter algebra, characterized by
this multiplicativity constraint. The connection between Baxter
algebras and the Riemann--Hilbert problem, which lurks in the
background here, remains largely unexplored, as of today.

Renormalization in the MS scheme can now be summarized in one
sentence: with the character $\phi$ given by the Feynman rules in
a suitable regularization scheme and well-defined on any small
curve around $\ve=0$, find the Birkhoff decomposition
$\phi_+(\ve)=\phi_-\star\phi$.

 The unrenormalized analytic
expression for a graph $\Gamma$ is then $\phi[\Gamma](\ve)$, the
MS-counterterm is $S_{{ MS}}(\Gamma)\equiv\phi_-[\Gamma](\ve)$ and
the renormalized expression is the evaluation $\phi_+[\Gamma](0)$.
Once more, note that the whole Hopf algebra structure of Feynman
graphs is present in this group: the group law demands the
application of the coproduct, $\phi_+=\phi_-\star \phi\equiv
S_{MS}^\phi\star\phi$.

But still, one might wonder what a huge group this group of
characters really is. What one confronts in QFT is the group of
diffeomorphisms of physical parameter: low and behold, changes of
scales and renormalization schemes are just such (formal)
diffeomorphisms. So, for the case of a massless theory with one
coupling constant $g$, for example, this just boils down to formal
diffeomorphisms of the form $$g\to \psi(g)=g+c_2 g^2+\ldots.$$ The
group of one-dimensional diffeomorphisms of this form looks much
more manageable than the group of characters of the Hopf algebras
of Feynman graphs of such a theory.

\subsection{Diffeomorphisms of physical parameters}
Thus, it would be very nice if the whole Birkhoff decomposition
could be obtained at the level of diffeomorphisms of the coupling
constants. The crucial ingredient is to realize the role of a
standard QFT formula of the form \be g_{\rm new}=g_{\rm old}\;Z_1
Z_2^{-3/2},\ee  which expresses how to obtain the new coupling in
terms of a diffeomorphism of the old. This was achieved in
\cite{RHII}, recognizing this formula as a Hopf algebra
homomorphism from the Hopf algebra of diffeomorphism to the Hopf
algebra of Feynman graphs, regarding $Z_g=Z_1/Z_2^{3/2}$, a series
over counterterms for all 0PI graphs with the external leg
structure corresponding to the coupling $g$, in two different
ways. It is at the same time  a formal diffeomorphism in the
coupling constant $g_{\rm old}$ and a formal series in Feynman
graphs. As a consequence, there are two competing coproducts
acting on $Z_g$. That both give the same result defines the
required homomorphism, which transposes to a homomorphism from the
largely unknown group of characters of ${\cal H}$ to the
one-dimensional diffeomorphisms of this coupling.

The crucial fact in this is the recognition of the Hopf algebra
structure of diffeomorphisms by Connes and Moscovici \cite{CM}:
Assume you have formal diffeomorphisms $\phi,\psi$ in a single
variable \be x\to\phi(x)=x+\sum_{k>1}c_k^\phi x^k,\label{expa}\ee
and similarly for $\psi$. How do you compute the Taylor
coefficients $c^{\phi\circ\psi}_k$ for the composition
$\phi\circ\psi$ from the knowledge of the Taylor coefficients
$c_k^\phi,c_k^\psi$? It turns out that it is best to consider the
Taylor coefficients
\be\delta_k^\phi=\log(\phi^\prime(x))^{(k)}(0)\label{CMT}\ee
instead, which are as good to recover $\phi$ as the usual Taylor
coefficients. The answer lies then in a Hopf algebra structure:
\be\delta^{\phi\circ\psi}_k=m\circ(\tilde{
\psi}\otimes\tilde{\phi} )\circ\Delta_{CM}(\delta_k), \ee where
$\tilde{\phi},\tilde{\psi}$ are characters on a certain Hopf
algebra ${\cal H}_{CM}$ (with coproduct $\Delta_{CM}$) so that
$\tilde{\phi}( \delta_i)$ $=$ $\delta_i^\phi$, and similarly for
$\tilde{\psi}$. Thus one finds a Hopf algebra with abstract
generators $\delta_n$ such that it introduces a convolution
product on characters evaluating to the Taylor coefficients
$\delta_n^\phi,\delta_n^\psi$, such that the natural group
structure of these characters agrees with the diffeomorphism
group.

It turns out that this Hopf algebra of Connes and Moscovici is
intimately related to rooted trees in its own right \cite{CK1},
signalled by the fact that it is linear in generators on the rhs,
as are the coproducts of rooted trees and graphs.\footnote{Taking
the $\delta_n$ as naturally grown linear combination of rooted
trees imbeds the commutative part of the Connes-Moscovici Hopf
algebra in the Hopf algebra of rooted trees, which on the other
hand allows for extensions similar to the ones needed by Connes
and  Moscovici. Details are in \cite{CK1}, with an extended
discussion of generalized natural growth and bicrossed product
structures in \cite{CKnew}.}

 There
are a couple of basic facts which enable to make in general the
transition from this rather foreign territory of the abstract
group of characters of a Hopf algebra of Feynman graphs (which, by
the way, equals the Lie group assigned to the Lie algebra with
universal enveloping algebra the dual of this Hopf algebra) to the
rather concrete group of diffeomorphisms of physical observables.
These steps are \begin{itemize}
\item Recognize that $Z$ factors are given as counterterms over
formal series of graphs starting with 1, graded by powers of the
coupling, hence invertible.
\item Recognize  the series $Z_g$ as a formal diffeomorphism,
with Hopf algebra coefficients.
\item Establish that the two competing Hopf algebra structures of
diffeomorphisms and graphs are consistent in the sense of a Hopf
algebra homomorphism.
\item Show that this homomorphism transposes to a Lie algebra and
hence Lie group homomorphism.
\end{itemize}
This works out extremely nicely, with details given in
\cite{RHII}. In particular, the effective coupling $g_{\rm
eff}(\ve)$ now allows for a Birkhoff decomposition in the space of
formal diffeomorphisms \be g_{\rm eff}(\ve)=g_{\rm eff
-}(\ve)^{-1}\circ g_{\rm eff +}(\ve) \ee where $g_{\rm eff
-}(\ve)$ is the bare coupling and $g_{\rm eff +}(0)$ the
renormalized effective coupling while the coupling can be regarded
as the coupling in any theory which has a single interaction term.
If there are multiple interaction terms in the Lagrangian, one
finds similar results relating the group of characters of the
corresponding Hopf algebra to the group of formal diffeomorphisms
in the multidimensional space of coupling constants.

There are some false but pertinent claims in the literature with
regards to the extension of this result to massless QED and other
theories with spin \cite{Brou}. The above results hold as they
stand for massless QED, with the relevant Hopf algebra
homomorphism given by $e_{\rm new}=Z_3^{-1/2}e_{\rm old}$.

The confusion arises from the assumption that in theories with
spin, the noncommutativity of Green functions demands the use of a
noncommutative noncocommutative Hopf algebra, ignoring the
commutative Hopf algebra obtained from the pre-Lie algebra of
graphs insertions above. This is patently wrong and ignores the
fact that the relevant characters on the Hopf algebra are given by
the coefficient functions of the tree-level form factors, and
these coefficient functions are scalar characters \be
\rho(\Gamma_1\Gamma_2)=\rho(\Gamma_2\Gamma_1)=\rho(\Gamma_1)\rho(\Gamma_2),\ee
on a commutative Hopf algebra, with all noncommutativity residing
in the matrix structures multiplying those form-factors. A typical
such character $\rho$ is provided by the Feynman rules $\phi$ of
massless QED applied to a fermion self-energy graph $\Gamma$ with
external momentum $\pslash$  say, where we have \be \phi(\Gamma)=
\rho(\Gamma)\pslash,\ee with
$\rho(\Gamma)=Tr(\pslash\phi(\Gamma))/p^2$, and $\pslash=\phi({\bf
res}(\Gamma))$, and indeed, the matrix structure of $\pslash$
plays no further role with respect to the commutativity of ${\cal
H}$.

\section{Multiple rescalings} So we have singled out $MS$ as a special character,
providing us with a  Birkhoff decomposition. So what about other
renormalization schemes: are they less meaningful? Not quite.

Let us come back to unrenormalized Feynman graphs, and their
evaluation by some chosen character $\phi$, and let us also choose
a renormalization scheme $R$. The group structure of such
characters on the Hopf algebra can be used in an obvious manner to
describe the change of renormalization schemes. This has very much
the structure of a generalization of Chen's Lemma \cite{Chen}.
\subsection{Chen's Lemma}
Consider $S_R\star\phi$. Let us change the renormalization scheme
from $R$ to $R^\prime$. How is the renormalized character
$S_{R^\prime}\star\phi$ related to the renormalized character
$S_R\star\phi$? The answer lies in the group structure of
characters: \be S_{R^\prime}\star\phi=[S_{R^\prime}\star S_R\circ
S]\star [S_R\star\phi],\ee which generalizes Chen's Lemma on
iterated integrals \cite{Chen}. We inserted a unit $\eta$ with
respect to the $\star$-product in form of $\eta=S_R\circ S\star
S_R\equiv S_R^{-1}\star S_R$, and can now read the
rerenormalization, switching between the two renormalization
schemes, as composition with the renormalized character
$S_{R^\prime}\star S_R^{-1}$. Note that $S_{R^\prime}\star
S_R\circ S$ is a renormalized character indeed: if $R,R^\prime$
are both self-maps of $V$ which do not alter the short-distance
singularities as discussed before, then in the ratio
$S_{R^\prime}\star S_R\circ S$ those singularities drop out.

Similar considerations apply to a change of scales which determine
a character \cite{Chen}. If $\mu$ is a dimensionful parameter
which dominates the process under consideration and which appears
in a character $\phi=\phi(\mu)$, then the transition
$\mu\to\mu^\prime$ is implemented in the group by acting on the
right with the renormalized character
$\psi_{\mu,\mu^\prime}^\phi:=\phi(\mu)\circ
S\star\phi(\mu^\prime)$ on $\phi(\mu)$,
\be\phi(\mu^\prime)=\phi(\mu)\star \psi_{\mu,\mu^\prime}^\phi.\ee
Let us note that this Hopf algebra structure can be efficiently
automated as an algorithm for practical calculations  exhibiting
the full power of this combinatorics \cite{BK1}.

Now, assume we compute Feynman graphs by some Feynman rules in a
given theory and decide to subtract UV singularities at a chosen
renormalization point $\mu$. This amounts, in our language, to
saying that the map $S_R$ is parametrized by this renormalization
point: $S_R=S_{R_\mu}$. Typically, one has for $R_\mu:V\to V$, \be
R_\mu(ab)=R_\mu(a)\,R_\mu(b),\label{ch} \ee in accordance with but
stronger than the multiplicativity constraints.

Then, let $\Phi(\mu,\mu^\prime)$ be the ratio
$\Phi(\mu,\mu^\prime)=S_{R_\mu}\star\phi(\mu^\prime)$. We then
have the groupoid law as part of the before-mentioned Chen's lemma
\cite{Chen,review} \be
\Phi(\mu,\eta)\star\Phi(\eta,\mu^\prime)=\Phi(\mu,\mu^\prime),\ee
thanks to Eq.(\ref{ch}).

\subsection{Automorphisms of ${\cal H}$}
We further note that the use of external structures always allows
to pullback renormalization schemes to automorphisms of the Hopf
algebra by solving the equation \be S_R^\phi\circ S=\phi\circ
\Theta_R, \ee for the automorphism $\Theta_R:{\cal H}\to{\cal H}$.
Starting from primitive elements $\Gamma$, this can be solved to
determine appropriate external structures, following the guidance
of \cite{Chen}, solving by a recursion over the bidegree.

The full group structure of the group of characters of the Hopf
algebra has barely been used in practice yet, with some notable
exceptions in \cite{BK1,BK}, but it provides an enormously rich
set of new tools for the investigation of QFT. One striking aspect
is that it allows insight into the hardest problem of QFT:
understanding the analytic aspects of the perturbative expansion,
by laying bare the way in which analytic input enters the
combinatorics of renormalization, and by allowing nonperturbative
results completely based on the Hopf and Lie algebra structures,
for example the beautiful duality between the scaling variable and
the coupling discovered in \cite{BK4}.

Harvesting the combinatorial structure of QFT emphasizes the
crucial role played by bidegree one graphs and their residues.
Relations between such diagrams  are often consequences of
symmetries in the Lagrangian -BRST symmetry after the quantization
of a local gauge symmetry, supersymmetry, or both. Even more
striking though are relations which can not be traced back to such
an origin. Typically they come as analytic surprises after the
calculation of different diagrams. To get an idea of an underlying
conceptual source for such relations, we have to look at our main
theme, insertion and elimination of subgraphs, more closely.

\section{Derivations on the Hopf algebra}
Having defined the Hopf- and Lie algebras of Feynman graphs, it is
profitable to look into representations of the Lie algebra as
derivations on the Hopf algebra, following \cite{CKnew}.

\subsection{Representations of ${\cal L}$}
The Lie algebra ${\cal L}$ gives rise to two representations
acting as derivations on the Hopf algebra ${\cal H}$: \be
Z^+_\Gamma\times \delta_X=\delta_{X\star\Gamma}\ee and \be
Z^-_\Gamma\times\delta_X=\sum_i \langle
Z^+_\Gamma,\delta_{X^\prime_{(i)}}\rangle
\delta_{{X^{\prime\prime}_{(i)}}}.\ee Furthermore, we stress again
that any term in the coproduct of a 1PI graph $\Gamma$ determines
gluing data $G_i$ such that
$$\Gamma=\Gamma^{\prime\prime}_{(i)}\star_{G_i}\Gamma^\prime_{(i)},
\forall i.$$ Here, $G_i$ specifies vertices in
$\Gamma^{\prime\prime}_{(i)}$ and bijections of their types with
the elements of $\Gamma^{\prime}_{(i)}$ such that $\Gamma$ is
regained from its parts:
$$\pica .$$
The first line gives a term $(i)$ in the coproduct, decomposing
this graph into its only divergent subgraph (assuming we have
chosen $\phi^3$ in six dimensions, say) and the corresponding
cograph, the second line shows the gluing  $G_i$ for this term, in
this example.

\noindent We want to understand the commutator \be
[Z_{\Gamma_1}^+,Z_{\Gamma_2}^-] ,\ee acting as a derivation on the
Hopf algebra element $\delta_X$. To this end introduce \be
Z_{[\Gamma_1,\Gamma_2]}\times\delta_X=\sum_i \langle
Z_{\Gamma_2}^+,\delta_{X^\prime_{(i)}}\rangle
\delta_{X^{\prime\prime}_{(i)}\star_{G_i}\Gamma_1}.\ee Here, the
gluing operation $G_i$ still acts such that each topologically
different graph is generated with unit multiplicity:
$$\picb .$$

\subsection{Insertion and Elimination}
So finally we are free to exchange graphs, by eliminating one and
inserting the other. \begin{theorem} \cite{CKnew}.  For all 1PI
graphs $\Gamma_i$, s.t.~${\bf res}(\Gamma_1)={\bf res}(\Gamma_2)$
and ${\bf res}(\Gamma_3)={\bf res}(\Gamma_4)$, the bracket
\begin{eqnarray*}
[Z_{[\Gamma_1,\Gamma_2]},Z_{[\Gamma_3,\Gamma_4]}] & = &
+Z_{[\overline{Z_{[\Gamma_1,\Gamma_2]}\times\delta_{\Gamma_3}},\Gamma_4]}
-Z_{[\Gamma_3,\overline{Z_{[\Gamma_2,\Gamma_1]}\times\delta_{\Gamma_4}}]}\\
& & -
Z_{[\overline{Z_{[\Gamma_3,\Gamma_4]}\times\delta_{\Gamma_1}},\Gamma_2]}
+Z_{[\Gamma_1,\overline{Z_{[\Gamma_4,\Gamma_3]}\times\delta_{\Gamma_2}}]}\\
 & &
 -\delta^K_{\Gamma_2,\Gamma_3}Z_{[\Gamma_1,\Gamma_4]}+\delta^K_{\Gamma_1,\Gamma_4}Z_{[\Gamma_3,\Gamma_2]},
\end{eqnarray*}
defines a Lie algebra of derivations acting on the Hopf algebra
${\cal H}$ via \be Z_{[\Gamma_i,\Gamma_j]}\times
\delta_X=\sum_I\langle Z^+_{\Gamma_2},\delta_{X^\prime_{(i)}}
\rangle \delta_{X^{\prime\prime}_{(i)}\star_{G_i}\Gamma_1},\ee
where the gluing data $G_i$ are normalized as before.\end{theorem}

The Kronecker $\delta^K$ terms just eliminate the overcounting
when combining all cases in a single equation.

We note that $Z_{[\Gamma,\Gamma]}\times \delta_X=k_\Gamma
\delta_X$, where $k_\Gamma$ is the number of appearances of
$\Gamma$ in $X$ and where we say that a graph $\Gamma$ appears $k$
times in $X$ if $k$ is the largest integer such that \be \langle
\Gamma^k\otimes {\rm id},\Delta(\delta_X)\rangle\ee is
non-vanishing. Also $I: Z_{[\Gamma_1,\Gamma_2]}\to
Z_{[\Gamma_2,\Gamma_1]}$ is an anti-involution such that \be
I([Z_{[\Gamma_1,\Gamma_2]},Z_{[\Gamma_3,\Gamma_4]}])=
-[I(Z_{[\Gamma_1,\Gamma_2]}),I(Z_{[\Gamma_3,\Gamma_4]})].\ee
Furthermore
$$[Z_{[\Gamma_1,\Gamma_2]},Z_{[\Gamma_2,\Gamma_1]}]=Z_{[\Gamma_1,\Gamma_1]}-Z_{[\Gamma_2,\Gamma_2]}.$$

\noindent By construction, we have
\begin{prop}
$$Z^+_\Gamma\equiv Z_{[\Gamma,{\bf res}(\Gamma)]},$$ $$Z^-_\Gamma\equiv
Z_{[{\bf res}(\Gamma),\Gamma]},$$  and
$[Z^-_X,Z^-_Y]=-Z^-_{\overline{[Z^+_X,Z^+_Y]}}.$\end{prop}

Finally, \begin{prop}
\begin{eqnarray*}
[Z_{[\Gamma_1,{\bf res}(\Gamma_1)]},Z_{[{\bf
res}(\Gamma_2),\Gamma_2]}] & = &
 \delta^K_{{\bf res}(\Gamma_1),{\bf res}(\Gamma_2)}Z_{[\Gamma_1,\Gamma_2]}+\delta^K_{\Gamma_1,\Gamma_2}
 Z_{[{\bf res}(\Gamma_2),{\bf res}(\Gamma_1)]}\\
   & & - Z^-_{\overline{Z_{[{\bf res}(\Gamma_1),\Gamma_1]}\times\delta_{\Gamma_2}}}
   - Z^+_{\overline{Z_{[{\bf
res}(\Gamma_2),\Gamma2]}\times\delta_{\Gamma_1}}}.
\end{eqnarray*}\end{prop}

These derivations provide a convenient mean to relate the
insertion of subgraphs at different places to Galois symmetries in
Feynman graphs, to be discussed below.

\section{Primitivity}
The letters in which QFT wants to be formulated are primitive 1PI
graphs. But QFT speaks in manners more subtle than words: the
letters do not come in linear order, but are inserted into each
other in a highly structured way as we saw. How fluent are we in
that language? Well, the combinatorial structures revealed so far
allow us to decipher the content of QFT -the general Feynman
graphs- completely in terms of these underlying letters. This
features the residue as the central notion in quantum field
theory, with each letter providing its own unique residue, a
renormalization group invariant which connects the topology of the
graph to number theory \cite{Book,BKold1,BKold2,BBB,DK4}. Let us
discuss the disentanglement of Feynman graphs into these residues
now.

Consider $\Gamma_1\star_i\Gamma_2$, for primitive graphs
$\Gamma_1,\Gamma_2$ and insertion at a compatible vertex $i$. Both
graphs provide a first order pole,  how do these two first order
poles  determine the second order pole in
$\Gamma_1\star_i\Gamma_2$? And how are the higher order poles
determined in general? Such questions can now be answered
completely thanks to the knowledge of the Hopf algebra structure.
\subsection{Higher Poles}
The explicit formulas in \cite{RHII} allow to find the
combinations of primitive graphs into which higher order poles
resolve. The weights are essentially given by iterated integrals
which produce coefficients which generalize the tree-factorials
obtained for the undecorated Hopf algebra in \cite{Chen,KD,BK1}.
Iterated application of this formula allows to express inversely
the first-order poles contributing to the $\beta$-function as
polynomials in Feynman graphs free of higher-order poles.

This decomposition into pole terms is accompanied by a
decomposition according to the bidegree \cite{BK3,Foissy}. In
practice, this decomposition is evident for subgraphs with two
legs in a massless theory, using that the only effect of
self-energies is to raise edge variables to non-integer scaling
degrees in Eq.(\ref{int}) \cite{Book}. We thus concentrate on the
decomposition of vertex subgraphs.

It is very instructive to use automorphisms $\Theta:{\cal H}\to
{\cal H}$ of the Hopf algebra of specified graphs which vary
external structures. We will assume that $\Theta$ is chosen in a
way such that $R_\Theta\circ\phi:=\phi\circ\Theta$ defines a map
$R_\Theta:V\to V$ in accordance with the usual requirements we
impose on renormalization maps. Typically, $\Theta$ will pose
further conditions on momenta attached to external edges of vertex
correction subgraphs and thus modifies external structures so that
an appropriate representation of Feynman graphs will be by
suitably colored graphs. Let then \be S_\Theta[(\Gamma,\sigma)]
=-\Theta\left[(\Gamma,\sigma)+\sum_{\gamma\subset\Gamma}(\Gamma/\gamma\star_{G_i}S_\Theta[(\gamma,\sigma)])\right],
\ee with gluing data $G_i$ as before and where we let
$\Theta[(\Gamma,\sigma)]:=(\Gamma,\Theta(\sigma))$ be such a
change of external structure. If we choose $\Theta$ to be trivial
on graphs with two external legs, but nontrivial on interaction
graphs (setting, for example, all but two chosen external momenta
to zero), one gets
\begin{theorem}
${\rm bid}(S_\Theta\star {\rm id}((\Gamma,\sigma)))=0$ and ${\rm
bid}(S_\Theta((\Gamma,\sigma)))=1$.
\end{theorem}
This is indeed obvious when we compare with the principle of
multiplicative subtraction: in the above: recursively, each
(sub)graph is decomposed into the image of ${\rm id}-\Theta$ or
$\Theta$. The usual recursion over subgraphs then gives a complete
decomposition over the bidegree such that each superficial
divergent subgraph is in the image of $\Theta$. Hence,
\begin{cor}
For each $\Theta$ as above, $\Gamma=\sum_{i=0}^{{\rm
bid}(\Gamma)}\Gamma_i$, with ${\rm bid}(\Gamma_i)=i$ and
$\Gamma_i=\prod_{j=1}^i\gamma_j$, with ${\rm bid}(\gamma_j)=1$,
\end{cor}
a complete factorization of higher bidegree graphs into primitive
ones: we get indeed a decomposition of $\Gamma$ in terms of
increasing bidegree such that all subgraphs in the term of highest
bidegree are in the image of $\Theta$. A simple analytic argument
based on the universal integral representation Eq.(\ref{int}) then
gives the higher order pole terms in terms of residues. As an
example, we get for the  graph $\Gamma=\Gamma_0\star_i\gamma$ of
the introduction, with ${\rm bid}(\Gamma)=2$: {\large \be
S_\Theta\star {\rm id}(\Gamma)=
\left(\vv,\sigma\right)(p_1,p_2)-\left(\vv,\Theta(\sigma)\right)(p_1)$$
$$ -\left(\vvx,\sigma\right)(p_1,p_2)
+\left(\vvx,\Theta(\sigma)\right)(p_1),\label{eeee}\ee} where we
let $\sigma$ be an evaluation at external momenta
$p_1,p_1+p_2,p_2$ and $\Theta(\sigma)$ be an evaluation at zero
momentum transfer, keeping $p_1$. Upon evaluation by $\phi$, in
the second line, the subgraph in a circle is inserted as a
subintegral into $\phi(\Gamma_0)$ such that it also has zero
momentum transfer at the appropriate leg. It modifies the integral
corresponding to $\phi(\Gamma_0)$ only in the dependence on a
single edge variable corresponds to the distinguished momentum
upon which the subgraph still depends when evaluated with the
external structure $\Theta(\sigma)$. This  results in the same
factorization in calculations as one gets from self-energy graphs.

It is easy to see that the four terms above combine to an Feynman
integral
$$\phi\left(S_\Theta\star {\rm id}(\Gamma)\right) $$ which is
convergent, and that the first and the third, as well as the
second and the fourth term in Eq.(\ref{eeee}), combine to a
Feynman integral free of subdivergences, while the fourth term
alone is of bidegree two, but decomposes in the requested manner,
due to the fact that the insertion of the subintegral only
modifies the dependence on a single edge variable in
$\phi(\Gamma_0)$. Note that this decomposition amounts to adding
zero in a way such that each subdivergence has the desired
external structure.

\subsection{The scattering type formula}
All this combines nicely to a scattering type formula, explicitly
worked out for the case of $\phi^3$ in \cite{RHII}. Underlying are
some asymptotic scaling properties of graded complex Lie groups
worked out in \cite{RHII} as well. It automatically reproduces the
right weights with which residues combine to coefficients of
higher pole terms, taking into account both the grading by loop
number and the bidegree.

In this context, locality of counterterms ensures that a
modification of scales will not change the negative part of the
Birkhoff decomposition of the character $\g$ of ${\cal H}$ under
consideration (following the notation of \cite{RHI}, we let $\g$
be the evaluation of the unrenormalized Feynman rules $\phi$ on an
infinitesimal circle around $\ve=0$). Hence,
$$ \g_- (\ve) \, \t_{t \ve} (\g_- (\ve)^{-1}) \ \hbox{is
convergent for} \ \ve \ra 0 \, ,
$$ where $\t_{t\ve}$ implements a one-parameter group of rescalings
\cite{RHI}.  The generator $\b = \left( \frac{\partial}{\partial
t} \, F_t \right)_{t=0}$ of this one parameter group is related to
the {\it residue} of $\g$
$$ \build{\rm Res}_{\ve = 0}^{} \g = - \left(
\frac{\partial}{\partial u} \, \g_- \left( \frac{1}{u} \right)
\right)_{u=0}  $$ by the simple equation, \be \b = Y \, {\rm Res}
\, \g \, ,  \ee where $Y = \left( \frac{\partial}{\partial t} \,
\t_t \right)_{t=0}$ is the grading.

Amazingly, one can give $\g_- (\ve)$ in closed form as a function
of $\b$ \cite{RHI}. We shall for convenience introduce an
additional generator in the Lie algebra of $G$ (i.e.~primitive
elements of $\Hc^*$) such that,
$$ [Z_0 , X] = Y(X) \qquad \fl \, X \in \hbox{Lie} \ G \, ,$$
where $Y$ implements the grading by the loop number. The
scattering type formula for $\g_- (\ve)$ is then, \be \g_- (\ve) =
\lim_{t \ra \ify} e^{-t \left( \frac{\b}{\ve} + Z_0 \right)} \,
e^{t Z_0} \, , \ee  exemplifying
 that the higher pole structure of the divergences is uniquely
determined by the residue.

Note that the above decomposition into residues allows for a
systematic investigation into the properties of the insertion
operad underlying the insertion of subgraphs.

This operad has as indexed composition $\circ_{v,b}$ for the
insertion of a subgraph $\gamma$, at a chosen place $v$ (an edge
or vertex) in another graph $\Gamma$, using a chosen bijection $b$
of $\gamma^{[1]}_{\rm int}$ with $f_v$. Relabelling rules of
internal edges and vertices such that axioms of an operad are
fulfilled are straightforward \cite{review}. We can now start
asking for the question how a variation of $\circ_{v,b}\to
\circ_{v^\prime,b^\prime}$ modifies the higher order pole terms,
with invariance of the highest degree pole terms easily obtained
from the scattering formula, as well as its decomposition into the
residues  of the underlying bidegree one graphs.

Invariances of the residue of a higher bidegree graph reflect, as
we will see, number-theoretic properties of a graph and will be
discussed below. But let us first comment on the usefulness of
Hochschild cohomology in the renormalization process, which will
finish our combinatorial review of standard renormalization
theory.

\section{Renormalization and Hochschild Cohomology}
A particularly nice way to proof locality of counterterms and
finiteness of renormalized Green functions can be obtained using
the Hochschild properties of the operator $B_+^x$, where $x$
indicates an appropriate primitive graph and its gluing data.
Indeed it raises the bidegree by one unit and is therefore a
natural candidate to obtain such feasts. To do this nicely, it
pays to describe how to map Feynman graphs to decorated rooted
trees.
\subsection{From Graphs to Trees}
Let us set out to define a Hopf algebra of non-planar decorated
rooted trees (see \cite{Foissy} for a census of such algebras)
${\cal H}_{DRT}$ and a homomorphism $\rho$ from the one of graphs
${\cal H}_{FG}$ to this one, $\rho: {\cal H}_{FG}\to {\cal
H}_{DRT}$, \be
\rho(\Gamma_1\Gamma_2)=\rho(\Gamma_1)\rho(\Gamma_2)\ee such that
\be \left[\rho\otimes\rho\right]
\circ\Delta_{FG}(\Gamma)=\Delta_T(\rho(\Gamma)).\ee Here,
$\Delta_T$ and $\Delta_{FG}$ are the coproducts in the Hopf
algebras of rooted trees and Feynman graphs. The range of $\rho$
defines a sub-Hopf algebra ${\cal H}_\rho\subset {\cal H}_{DRT}$,
and $\rho$ is a bijection between ${\cal H}_{FG}$ and this closed
Hopf algebra ${\cal H}_\rho$. We let $1$ be the unit in both Hopf
algebras so that $\rho(1)=1$ and identify scalars.

In both algebras we have a decomposition with respect to the
bidegree \be {\cal H}_{FG}=\oplus_{i\geq 0} {\cal
H}_{FG}^{(i)}\;\mbox{and}\;{\cal H}_{DRT}=\oplus_{i\geq 0} {\cal
H}_{DRT}^{(i)}.\ee We set, to start an induction over the
bidegree, ${\cal H}_{FG}^{(i)}={\cal H}_{DRT}^{(i)},\; i=0,1$,
where we identify a primitive graph $\Gamma$ with the decorated
rooted tree $\rho(\Gamma)=(\ast,\Gamma)\in {\cal H}_{DRT}^{(1)}$.

Then, for some positive integers $r$ and $k_j$, $j=1,\ldots,r$,
any non-primitive 1PI graph $\Gamma$ can be written at most in $r$
different  forms
$$\Gamma=\prod_{i=1}^{k_j} \Gamma_j  \star_{j,i} \gamma_{j,i},\;\forall j=1,\ldots,r.$$
We call $\Gamma$ overlapping if $r>1$. We saw an example for the
case  $r=2$ earlier, in Fig.(\ref{f1}).  Then, we set \be
\rho(\Gamma) = \sum_{j=1}^r B_+^{\Gamma_j,G_{j,i}}\left[
\prod_{i=1}^{k_j}\rho(\gamma_{j,i})\right],\ee with appropriate
gluing data $G_{j,i}$, obtained from $\Delta_{FG}$ as before, see
Eq.(\ref{glue}). One immediately proves by induction the required
properties of $\rho$, using \be
\Delta_T(\rho(\Gamma))=\rho(\Gamma)\otimes 1+\sum_{j=1}^r
(1\otimes
B_+^{\Gamma_j,G_{j,i}})\Delta_T(\prod_{i=1}^{k_j}\rho(\gamma_{j,i}))
,\ee which allows for an inductive construction of $\rho$. The so
obtained decorated rooted trees will have decorated vertices where
the decorations consist of bidegree one (primitively divergent)
graphs and gluing data, which store information how to glue the
descendent branches into that decoration. Note that the action of
the $B_+$ operator is well defined under the coproduct: if
$X=B_+^{\Gamma,G_Y}(Y)$, then $B_+^{\Gamma,G_Y}(Y^{\prime\prime})$
is well-defined, as the gluing data only use the external legs of
$Y$, and ${\bf res}(Y)={\bf res}(Y^{\prime\prime})$ for all
cographs $Y^{\prime\prime}$ appearing in the coproduct of $Y$. All
these operators $B_+^{\Gamma_j,G_{j,i}}$ are closed Hochschild
one-cocycles.

The map $\rho$ is not uniquely given though: the usual ambiguity
in a transition from a filtration to a degree means that in this
construction of $\rho$ in terms of increasing bidegree, $\rho$ can
always be modfied by terms of lower bidegree. By construction, we
fixed $\rho$ at bidegrees 0,1.  The remaining freedom is actually
an asset in practical calculations \cite{CK1,overl,BDK}.

\subsection{Locality and Finiteness}
By the above, we can restrict an inductive proof of locality of
counterterms and finiteness of renormalized Green functions to the
study of elements of the form $\Gamma=B_+^{\gamma_i,G_i}(X_i)\in
{\cal H}_{FG},$ where we use the same notation $B_+^{\gamma,G}$ in
the Hopf algebra of graphs as in the one of decorated rooted
trees.

We will proceed by an induction over the bidegree which is much
more natural than the usual induction over the number of loops.

Hence our task is: assume that $S_R\star\phi(\Gamma)$ is finite
and and $S_R(\Gamma)$ a local counterterm for all $\Gamma$ with
${\rm bid}(\Gamma)\leq k$. Show these properties for all $\Gamma$
with ${\rm bid}(\Gamma)=k+1$.

The start of the induction is easy: at unit bidegree,
$\phi(\Gamma)-R[\phi(\Gamma)]$ is finite and $S_R(\Gamma)$ is
local by assumption on $R$.

Let us assume we have established the desired properties of $S_R$
and $S_R\star\phi$ acting on all Hopf algebra elements up to
bidegree $k$. Assume ${\rm bid}(\Gamma)=k+1$. We have \be
\Gamma=B_+^{\gamma,G}(X),\ee where ${\rm bid}(\gamma)=1$, ${\rm
bid}(X)=k$, $X$ some monomial in the Hopf algebra.

Next, \be\Delta(\Gamma)=B_+^{\gamma,G}(X)\otimes 1+[1\otimes
B_+^{\gamma,G}]\Delta(X),\ee which expresses the fact that
$B_+^{\gamma,G}$ is a closed Hochschild one-cocycle.

Let us decompose, as usual,
$$\Delta(X)=X\otimes 1+1\otimes
X+X^\prime\otimes X^{\prime\prime}.$$ It is easy to see from the
structure of Eq.(\ref{int}) that we can define \be{\bf
B}_+(\phi;\phi;\gamma,G;X)\equiv\phi(B_+^{\gamma,G}(X))\ee and
extend this definition to a map ${\bf
B}_+(\phi;S_R\star\phi;\gamma,G;X)$ which glues the renormalized
results $S_R\star\phi$ into the integral $\phi(\gamma)$.

Using the Hochschild closedness of $B_+^{\gamma,G}$ one
immediately gets \be S_R\star\phi(\Gamma)=S_R(\Gamma)+{\bf
B}_+(\phi;S_R\star\phi;\gamma,G;X)\ee and \be S_R(\Gamma)= -R[{\bf
B}_+(\phi;S_R\star\phi;\gamma,G;X)].\ee From here, the induction
step boils down to a simple estimate using the fact that the
powercounting for asymptotically large internal loop momenta in
$\phi(\gamma)$ is modified by the insertion of $S_R\star\phi(X)$
(which is finite by assumption, having bidegree $k$) only by
powers of $\log(|k_e|)$, and that delivers the result easily, from
Eq.(\ref{int}).

\section{Unique Factorization and Dyson--Schwinger\\ Equations}
So far, we have described the combinatorial structures underlying
renormalization theory. As it befits a Hopf algebra, it all comes
down to the study of residues, invariants under diffeomorphisms of
continous quantum numbers, parametrizing the primitive elements of
the Hopf algebra. Locality hides in the fact that every primitive
element has a residue proportional to its graphical residue:
\be{\rm res}[\phi(\Gamma)]=\rho(\Gamma)\langle\sigma_P,\phi({\bf
res}(\Gamma) \rangle\,,\;\forall \Gamma, {\rm bid}(\Gamma)=1,\ee
where $\rho$ is a character on ${\cal H}$. This gives a
fascinating way of actually defining the tree level terms
$\langle\sigma_P,\phi({\bf res}(\Gamma))\rangle $ in the
Lagrangian as the residues of primitive graphs.

This, it occurs to me, is a proper way to define locality in
noncommutative QFT. For such theories the traditional notion of
locality is obscured by the fact that non-vanishing commutators of
spacetime coordinates exponentiate to non-local functions even at
the tree level \cite{Chep,Dou}. Not too surprisingly, counterterms
involve terms which are then non-local in the traditional sense of
not being powers of fields and their derivatives. Nevertheless,
the real question to my mind is if they obey the equation above:
are the residues (in the operator-theoretic sense) of the bidegree
one-part in the perturbative expansion the generators of the tree
level (the bidegree zero part) terms?

This is a well-known phenomenon  in the study of gauge theories
over ordinary spacetime \cite{inst}, which allows to recover the
Yang-Mills action from integrating the one-loop fermion
determinant. This has far reaching generalizations incorporating
the noncommutative case and connecting gauge theories to Connes'
spectral action, via an operator theoretic residue \cite{ColdF}.
The significance of such an approach certainly lies in the
emphasis it gives to the Dirac operator and the residue. But in
this, one-loop graphs played a distinguished role, and the most
promising way to generalize such results to the whole of QFT is to
factorize QFT completely into bidegree one graphs, into residues
and traces, that is.

So let us muse, in this final section, about the structure of a
perturbative expansion in general, using properties of the
Dyson--Schwinger equations. These quantum equations of motion are
equations which are formally solved in an infinite series of
graphs, providing a fixpoint for these equations, and evaluating
these graphs order by order using the renormalized character
$S_R\star\phi$. These equations come typically as integral, or
integro-differential equations with a firm reputation to be
unsolvable: indeed, their solution would solve and establish
quantum field theory in four dimensions.

We essentially learned above how to disentangle Feynman graphs
into primitive graphs of unit bidegree. How can we utilize this
fact in those equations?

A typical example of such an equation is \be\fx.\ee This equation
has as a fixpoint a formal series over graphs which starts like
\be \fy.\ee
\subsection{Unique Factorization into Residues}
To start a more systematic understanding of this equations let us
sketch here an approach based on the fact that the series of
graphs which provide a solution for it can be written with the
help of a commutative product which generalizes the shuffle
product (a rather interesting construct in its own right
\cite{MEH1}) of ordered sequences to our (partially ordered)
Feynman graphs.

To this end, let $w=(a_1,a_2,\ldots,a_n)$ be a sequence of
primitive graphs, a word, say, of letters ordered in such a
sequence.

We now can define a partial order on Feynman graphs by defining
the set $I(w)$ of all Feynman graphs compatible with the ordered
sequence $w$ as those 1PI graphs $\Gamma$ which fulfill \be
\langle Z_{a_1}\otimes\ldots\otimes
Z_{a_k},\Delta^{k-1}(\Gamma)\rangle =1.\ee Here, $Z_{a_j}$ are the
duals of the primitive elements $a_j$. Let
$\Sigma(w)=\sum_{\Gamma\in I(w)}\Gamma$ be the sum over all graphs
compatible with $w$. Note that with the underlying partial order
based on 'being a subgraph', \be
\Gamma_1\leq\Gamma_2\Leftrightarrow
Z^-_{\Gamma_1}\times\delta_{\Gamma_2}\not= 0,\ee we can write for
any graph $\Gamma$
\be\Delta(\Gamma)=\sum_{\Gamma_1,\Gamma_2\in{\cal
H}}\zeta(\Gamma_1,\Gamma_2)\Gamma_1\otimes\Gamma_2,\ee where
$\zeta(\Gamma_1,\Gamma_2)$ is the $\zeta$-function of the
incidence algebra assigned to this partial order \cite{Rota}.

The permutation group on $n$ elements acts naturally on the word
$w$ with $n$ letters. Let $n_\Gamma:=\sum_{\sigma\in
S_n}\sum_{\Gamma_1\in I(\sigma(w))}\langle Z_\Gamma,
\delta_{\Gamma_1}\rangle$. Then, we define $\overline{\Sigma}(w)$
to be the same sum as $\Gamma(w)$, but each graph with coefficient
$1/n_\Gamma$: \be\overline{\Sigma}(w)=\sum_{\Gamma\in
I(w)}\frac{1}{n_\Gamma}\Gamma.\ee

On words $w_1=(a_1,v_1),w_2=(b_1,v_2)$ ($a_1,b_1$ letters,
$v_1,v_2$ subwords) one has the usual commutative associative
product of rifle shuffles \be w_1\perp w_2= (a_1,v_1\perp
w_2)+(b_1,w_1\perp v_2).\ee Then, one can consider the commutative
associative product \be
\overline{\Sigma}(w_1)\cdot\overline{\Sigma}(w_2):=\overline{\Sigma}(w_1\perp
w_2).\ee The series of graphs giving a formal solution to the
Dyson Schwinger equation above  can now be obtained as the product
of two Euler factors
\be\frac{1}{1-\overline{\Sigma}(\gamma_1)}\cdot
\frac{1}{1-\overline{\Sigma}(\gamma_2)}.\ee Note that the fact
that the Dyson-Schwinger equation sums democratically over all
possible insertions of graphs into each other is crucial here.
Much more about this will be said in future work.

Not much stops us to consider actually an Euler product over all
primitive graphs to get a formal solution to Dyson--Schwinger
equations in general. We should just construct $\zeta$-functions
dedicated to a chosen Green function, defined via an Euler product
over primitive elements. Or does it?

First of all, we would like to have the structure of an Euler
product not only on the level of graphs, but would like it to be
compatible with the Feynman rules. That would deliver a genuine
factorization into residues, with far reaching consequences for
non-perturbative aspects, in particular with respect to the
notorious renormalon problem: indeed, if we had for example for
general Euler factors
$$ \phi\left(\frac{1}{1-\overline{\Sigma}(\gamma_1)}\cdot \frac{1}{1-\overline{\Sigma}(\gamma_2)}\right)=
\frac{1}{1-\phi(\overline{\Sigma}(\gamma_1))}
\frac{1}{1-\phi(\overline{\Sigma}(\gamma_2))},$$ we would easily
verify that  no chains of subgraphs $\gamma_2$ say would produce
the renormalon problem: factorization into Euler products resums
the perturbation series in a way eliminating the factorial growth
attached to renormalons, decomposing the whole perturbation series
into an Euler product over geometric series.

 Note that the existence of Feynman
rules compatible with the above factorization essentially would
establish a shuffle identity on Feynman graphs, in complete
generalization of the situation for iterated integrals. Again,
much more on that will be reported in future work.

A further difficulty in this approach is that a full D-S equation
mixes graphs with a varying number of external legs. In
particular, in the above example we avoided the presence of
overlapping divergences. In their study, to my mind, are hidden
some of the most valuable treasures of quantum field theory.

\subsection{Uniqueness of Factorization  and ideals in gauge theories}
In the presence of overlapping divergences, unique factorization
is lost. Reconsider Fig.(\ref{f1}), an example taken from
non-abelian gauge theory.
$$\fz$$
The linear combination $\Omega$ of the two graphs
$\Omega_1,\Omega_2$ in the first line describes either all ways of
inserting $\Gamma_1$ into $\omega_1$, or all ways of inserting
$\Gamma_2$ into $\omega_2$. $\Omega$ does not factorize in a
unique manner, a phenomenon typical for local gauge theories, as
will be demonstrated elsewhere. We only very briefly sketch some
relevant ideas here.

Comparing the situation with algebraic number theory (see
\cite{Stark} for an excellent review) gives the right idea how to
restore unique factorization: consider ideals. Apparently, if we
let $(a,b,c,\dots)$ the ideal of Feynman graphs containing letters
$(a,b,c,\dots)$, then we can restore unique factorization of the
ideal $(\Omega)$ as
\be(\Omega)=(\omega_1,\omega_2)(\omega_1,\Gamma_2)(\Gamma_1,\omega_2)(\Gamma_1,\Gamma_2)
.\ee This implies that we can build products of these components -
we must have
$$(\omega_1,\omega_2)(\omega_1,\Gamma_2)=(\omega_1\cdot\omega_1,\omega_1\cdot\Gamma_2,\omega_2\cdot\omega_1,\omega_2\cdot\Gamma_2)=(\omega_1),$$
$$(\Gamma_1,\omega_2)(\Gamma_1,\Gamma_2)=(\Gamma_1\cdot\Gamma_1,\Gamma_1\cdot\Gamma_2,\omega_2\cdot\Gamma_1,\omega_2\cdot\Gamma_2)=(\Gamma_1)$$
and
\be(\omega_1,\omega_2)(\Gamma_1,\omega_2)=(\omega_1\cdot\Gamma_1,\omega_1\cdot\omega_2,\omega_2\cdot\Gamma_1,\omega_2\cdot\omega_2)=(\omega_2),\ee
$$(\omega_1,\Gamma_2)(\Gamma_1,\Gamma_2)=(\omega_1\cdot\Gamma_1,\omega_1\cdot\Gamma_2,\Gamma_2\cdot\Gamma_1,\Gamma_2\cdot\Gamma_2)=(\Gamma_2).$$
But this demands that there are relations between those graphs:
for example, the product $\omega_2\cdot\Gamma_1$ is not allowed to
vanish, if we want to maintain the structure of an integral domain
for our democratic insertions of graphs. How can this be achieved
when for example there is no place $v$ in $\omega_2$ such that
${\bf res}(\Gamma_1)=f_v$?

The answer lies in the local gauge symmetry, which actually mixes
external structures. An easy calculation shows that the one-loop
transversal gauge boson propagator and the one-loop fermion
propagator are connected:
\be\rho_{F^2}\left(\fo\right)=\xi^{-1}\rho_{\bar{\psi}D\psi}\left(\fs\right)\frac{D-3}{D-1}Tr(I),\label{rel}\ee
where we define the  gauge parameter $\xi$ as the deviation from
the transversal gauge, that is we define the massless Euclidean
gauge boson propagator as $$
P^{-1}_{\mu\nu}(k)=\frac{g_{\mu\nu}}{k^2}+(\xi-1)\frac{k_\mu k_\nu
}{k^4},$$ and use suitable characters $\rho_P$. Note that both
sides of Eq.(\ref{rel}) are independent of $\xi$. This identity
alone is sufficient to guarantee that none of the above products
is degenerate. Eq.(\ref{rel}) allows for a very beautiful
reinterpretation of gauge symmetry which will be commented on in
much greater detail in future work.

As a final remark, note that along with such a factorization comes
a considerable simplification in the study of quantum action
principles and renormalization of gauge symmetries, as the Leibniz
rules for the relevant (BRST-) differentials ensure that one can
restrict the study of such questions to single primitively
divergent bidegree one factors, which allows to lift results as
\cite{Martin} to all orders once factorization is established.

\subsection{Galois symmetries in perturbative QFT}
In the above, we started drifting towards a treatment of Feynman
graphs as a ring, with associated field of fractions say, where
the role of primes is played by primitive graphs, and an Euler
product combined with an appropriate shuffle identity for Feynman
rules  should guide us towards an appropriate notion of a
$\zeta$-function for a given Green function.

It is essentially unique factorization which then summarizes all
the wisdom of the forest formulas with the renormalization group
playing the role of a group of units in that field, and with gauge
symmetries  making life interesting in the quest for unique
factorization.

But then, the solution of the Dyson-Schwinger equation can be
factorized in various different sets of primitives. This brings us
back to the behaviour of graphs which are build from similar
primitives, inserted into each other at different places though.
In the mental picture laid bare here this connects to differential
Galois theory: Consider the combination
$\Gamma_0(\star_i-\star_j)\gamma$, studied in the introduction,
Eq.(\ref{intro}).

We can consider the "differential equation" ($Z_{[{\bf
res}(\gamma),\gamma]}$ is a derivation!)  \be Z_{[{\bf
res}(\gamma),\gamma]}\times \delta_\Gamma=\Gamma_0,\ee which is
solved by the bidegree two graph $\Gamma=\Gamma_0\star_i\gamma$ as
well as by the bidegree two $\Gamma= \Gamma_0 \star_j\gamma$,
while the bidegree one primitive
$X=\Gamma_0(\star_i-\star_j)\gamma$ solves the homogenous equation
\be Z_{[{\bf res}(\gamma),\gamma]}\times \delta_X=0.\ee This
justifies to connect the insertion of subgraphs at various
different places with Galois symmetries, and was the motivation to
indeed look at invariants under such symmetries in Feynman graphs,
with a beautiful first result reported in \cite{BKK}: the
coefficient of the highest weight transcendental in the residues
of two graphs connected by such a symmetry is invariant. While
this is obvious, thanks to the scattering type formula, for the
coefficient of the highest pole in the regularization parameter,
it is a very subtle result for the residue in a graph of large
bidegree.

\section*{Conclusion and Acknowledgments}
Quantum field theory is to me still the most subtle construct we
came across so far in theoretical physics. In particular, in its
short distance singularities, it hides mathematical wisdom and
beauty which parallel its unmatched achievements as a predictive
theory. It is my hope that the recently discovered combinatorial
structures in quantum field theory provide powerful tools to
unravel these structures, for the benefit of the mathematician and
practitioner of quantum field theory alike.

Many thanks to Dennis Sullivan and the organizers of the {\em
Dennisfest} for giving me the opportunity to speak at this most
entertaining workshop.

It is a pleasure to thank Ron Donagi, Takashi Kimura and Jim
Stasheff for discussions on various algebraic and operadic
properties of Feynman graphs, and my friends and collaborators
David Broadhurst and Alain Connes for much needed help in
sharpening  more of the tools necessary to deal with QFT.

Partial support of this work by the Clay Mathematics Institute is
gratefully acknowledged.

\end{document}